\documentclass[12pt]{article}

\usepackage{fullpage}
\usepackage{multirow}
\usepackage{graphicx}
\usepackage{amsmath}
\usepackage{amssymb}
\usepackage{amsfonts}
\usepackage{natbib}
\usepackage{pstricks}
\usepackage{setspace} 
\usepackage{aeguill} 
\usepackage{psfig}
\usepackage{epsfig}

\begin{document} 

\title{\bf Global sensitivity analysis for models with spatially dependent outputs} 

\author{\bf Amandine MARREL$^1$, Bertrand IOOSS$^2$, \\
\bf Michel JULLIEN$^3$, Béatrice LAURENT$^4$ and Elena VOLKOVA$^5$}
\date{}

\maketitle 

\begin{center}

  $^1$ IFP, 92852 Rueil-Malmaison cedex, France
  
  $^2$ EDF R\&D, 6 Quai Watier, 78401 Chatou, France 
  
  $^3$ CEA, DEN, F-13108 Saint Paul lez
  Durance, France 

  $^4$ Institut de Mathématiques de Toulouse (UMR 5219), INSA de Toulouse, Universit\'e de Toulouse, France  

  $^5$ RRC ``Kurchatov Institute'', Institute of Nuclear Reactors, Russia
  
  \vspace{0.2cm}
  Corresponding author: B. Iooss ; Email: biooss@yahoo.fr\\
  Phone: +33 (0)1 30 87 79 69
  
  
  \vspace{0.5cm}
  
\end{center}

\doublespacing


\begin{abstract}
The global sensitivity analysis of a complex numerical model often calls for the estimation of variance-based importance measures, named Sobol' indices.
Metamodel-based techniques have been developed in order to replace the cpu time-expensive computer code with an inexpensive mathematical function, which predicts the computer code output.
The common metamodel-based sensitivity analysis methods are well-suited for computer codes with scalar outputs.
However, in the environmental domain, as in many areas of application, the numerical model outputs are often spatial maps, which may also vary with time. In this paper, we introduce an innovative method to obtain a spatial map of Sobol' indices with a minimal number of numerical model computations.
It is based upon the functional decomposition of the spatial output onto a wavelet basis and the metamodeling of the wavelet coefficients by the Gaussian process.
An analytical example is presented to clarify the various steps of our methodology.
This technique is then applied to a real hydrogeological case: for each model input variable, a spatial map of Sobol' indices is thus obtained.
\end{abstract}


\noindent {\bf Short title:} Spatial global sensitivity analysis



\section{INTRODUCTION}


Today, in different environments, there are sites with groundwater contaminated because of an inappropriate handling or disposal of hazardous materials or waste.
Such environmental or sanitary issues require the development of treatment or remediation strategies and, in all cases, a robust long-term prediction of behaviour. The indispensable simulation of global fluxes, such as water or pollutants, through the different environmental compartments involves many parameters.
 Numerical modeling is an efficient tool for an accurate prediction of the spreading of the contamination plume and an assessment of environmental risks associated to the site. 
However, it is well known that many input variables, such as hydrogeological parameters (permeabilities, porosities, etc.) or boundary and initial conditions (contaminant concentrations, aquifer level, etc.), are highly uncertain in the complex numerical models.
A systematic and exhaustive 3D characterization of sites is still impossible.

To deal with all these uncertainties, computer experiment methodologies based upon statistical techniques are useful.
For instance, we assume that $Y = f (\mathbf{X})$ is the real-valued output of a computer
code $f$. Its input variables are random and modeled by the random vector $\mathbf{X}=(X_1,\ldots,X_d) \in \mathcal{X}$, $\mathcal{X}$ being a bounded domain of $\mathbb{R}^d$, of known distribution.
The uncertainty analysis step is used to evaluate statistical parameters, confidence intervals or the density probability distribution of the model response \citep{derdev08}, while the global sensitivity analysis step is used to quantify the influence of the uncertainties of the model input variables (in their whole range of variations) on model responses \citep{salcha00}. Recent studies have applied different statistical methods of uncertainty and sensitivity analysis to environmental models \citep{hel93,nyccox98,fasesp03,volioo08,liltar09}.
All these methods have shown their efficiency in providing guidance to a better understanding of the modeling.

However, for the purpose of sensitivity analysis, four main difficulties can arise due to practical problems, especially when focusing on environmental risks: 
\begin{enumerate}
\item[P1)] physical models involve rather complex phenomena (they are non linear and subject to threshold effects) sometimes with strong interactions between physical variables
\item[P2)] computer codes are often too cpu time expensive to evaluate a model response, from several minutes to weeks
\item[P3)] numerical models take as inputs a large number of uncertain variables (typically $d>10$)
\item[P4)] the outputs of these numerical encompass many variables of interest, that can vary in space and time. 
\end{enumerate}

The first problem P1 is solved using variance-based measures. These ones can handle non-linear and non-monotonic relationships between inputs and output \citep{salcha00}.
These measures are based upon the functional ANOVA decomposition of any integrable function $f$ \citep{efrste81} and determine how to share of the variance of the output resulting from a variable $X_i$ or an interaction between variables \citep{sob93}:
\begin{equation}
  \label{eq:indordre1}
  S_i=\frac{\mbox{Var}\left[\mathbb{E}\left(Y|X_i\right)\right]}{\mbox{Var}(Y)} \;,
  \quad  S_{ij} = \frac{\mbox{Var}\left[\mathbb{E}\left(Y|X_i,
        X_j\right)\right]}{\mbox{Var}(Y)} - S_i -S_j \;,  \quad  S_{ijk} = \ldots
\end{equation} 
The interpretation of these coefficients, namely the Sobol' indices, is natural as all indices lie in $[0,1]$ and their
sum is one in the case of independant input variables. The larger the index value, the greater the importance of the variable related to
this index.
To express the overall output sensitivity to an input $X_i$, \citet{homsal96} introduce the total sensitivity index:
\begin{eqnarray}\label{eq:indtotal}
  S_{T_i} = S_i + \sum_{i<j} S_{ij} + \sum_{i<j<k} S_{ijk} + \ldots = \sum_{l \,\in \,\# i}S_l =1-\frac{\mbox{Var}\left[\mathbb{E}\left(Y|X_{\sim i}\right)\right]}{\mbox{Var}(Y)}
\end{eqnarray}
where $\# i$ represents all the ``non-ordered'' subsets of indices
containing index $i$ and $X_{\sim i}$ is the vector of all inputs except $X_i$. 
Thus, $\sum_{l \,\in \,\# i}S_l$ is the sum of
all the sensitivity indices with index containing $i$.

Unfortunately, the traditional or advanced Monte Carlo methods, which are used to estimate first order and total Sobol' indices, require a large number of model evaluations \citep{salann09}.
To overcome the problem P2, of too long a calculation time, and make uncertainty and sensitivity analysis tractable, various approaches based upon metamodeling were recently proposed \citep{koeowe96,klesar00,oakoha02}.
The key point consists of replacing the complex computer code by a mathematical approximation, called a metamodel, which is fitted from only a few experiments. The metamodel reproduces the behavior of the computer code in the domain of its influential parameters \citep{sacwel89,fanli06}.
Among all the metamodel-based solutions (polynomials, splines, neural networks, etc.), we focus our attention on the Gaussian process (Gp) model. It can be viewed as an extension of the kriging method, which is used for interpolating data in space \citep{chidel99}, to computer code data \citep{sacwel89,oakoha02}.
Many authors \citep[e.g. ][]{welbuc92,marioo07} have shown how the Gp model can be used as an efficient emulator of code responses, even in high dimensional cases (problem P3).

In this paper, we consider models subject to the four problems together (P1, P2, P3 and P4), which is an usual cas in model-based environmental studies.
We mainly pay attention to problem P4, that is the possible high dimension of model outputs.
In the application case studied in this paper, the costly numerical model yields spatial concentration maps.
These spatial outputs encompass several thousands of grid blocks, each with a concentration value.
This kind of problem cannot be tuned to a vectorial output problem because of its dimensionality: the metamodeling of this vectorial output cannot be solved referring to kriging or cokriging techniques \citep{fanli06}.
Therefore, we consider the model output as a functional output synthesized by its projection on an appropriate basis.
This problem of building a metamodel (based upon functional decomposition and Gp modeling) for a functional output has recently been addressed for one-dimensional outputs by \citet{shiwan07} and \citet{bayber07} and for two-dimensional outputs by \citet{higgat08}.

In the case of sensitivity analysis, a functional output is usually considered as a vectorial output and sensitivity indices relative to each input are computed for each discretized value of the output \citep{derdev08}.
To avoid the large amount of sensitivity index computations when applying such an approach, a few authors referred to various basis decompositions on the functional output, such as the principal component analysis \citep{cammck06,lammak09}.
Then, sensitivity indices are obtained for the coefficients of the expansion basis.

However, the full functional restitution of Sobol' indices remains an unexplored challenge.
In this paper, we propose an original and complete methodology to compute Sobol' indices at each location of the spatial output map.
Our approach consists of building a metamodel based upon wavelet decomposition as in \citet{bayber07} (restricted to the case of a temporal output).
This metamodel is then used to compute spatial Sobol' index maps (one map per input variable).
A map of the Sobol' index of an input $X_i$ shows the local and global influences of this input on the output. It can help to better understand the computer code results and can used to reduce more efficiently the uncertainties in the responses. Thus, to reduce the output variability at a given point of the map, we analyze all Sobol' maps and determine the most influential inputs.
Then, we can try to reduce the uncertainty of these inputs by accounting for additional measures. In addition, the global influence of each input over the whole space can be investigated to identify areas of influence and non-influence for this input.

Details about the Gp metamodel are given in the following section.
Then, a step by step description of our methodology is given in Section 3.
A synthetic test function is used to evidence the relevance of our choices and estimate the convergence of the algorithms.
Section 4 presents how our methodology is applied to a real environmental problem, which calls for the modeling of radionuclide groundwater migration (MARTHE code).
Then, a few points are discussed at the end of this paper.


\section{GAUSSIAN PROCESS METAMODELING}


This section introduces the Gp metamodel for the case of a single scalar output.
We consider $n$ realizations of a computer code. Each realization $y=f(\mathbf{x}) \in \mathbb{R}$ is an output of the computer code and corresponds to a $d$-dimensional input vector $\mathbf{x} = (x_1,\ldots,x_d) \in \mathcal{X}$. The $n$ points corresponding to the code runs are called the experimental design and are denoted as $X_s = ( \mathbf{x}^{(1)},\ldots,\mathbf{x}^{(n)} )$. The outputs are denoted as $Y_s = (y^{(1)},\ldots,y^{(n)})$ with $y^{(i)} = y(\mathbf{x}^{(i)})$ $\forall \; i=1..n$.
Gp modeling treats the deterministic response $y(\mathbf{x})$ as a realization of a random function $Y_{\mbox{\scriptsize Gp}}(\mathbf{x})$. This includes a regression part and a centered stochastic process \citep{sacwel89}. It can be written as:
\begin{equation}\label{eq:GP}
 Y_{\mbox{\scriptsize Gp}} ( \mathbf{x} ) = f_0 ( \mathbf{x} ) + Z ( \mathbf{x}) \;.
\end{equation}

The deterministic function $f_0(\mathbf{x})$ provides the mean approximation of the computer code. In our study, we use a one-degree polynomial model with $f_0(\mathbf{x})$ written as:
\[ f_0(\mathbf{x} ) = \beta_0 + \sum_{j = 1}^d \beta_j x_j \;, \]
where $\mbox{\boldmath $\beta$} = [ \beta_0, \ldots, \beta_k ]^t $ is the regression parameter vector.
It has been shown, for example in \cite{marsim05} and \cite{marioo07}, that such a function is sufficient, and sometimes necessary, to capture the global trend of the computer code.

The stochastic part $Z(\mathbf{x})$ is a Gaussian centered process fully characterized by its covariance function:
$\mbox{Cov} ( Z ( \mathbf{x} ), Z ( \mathbf{u} ) ) = \sigma^2 R ( \mathbf{x}, \mathbf{u} ),$
where $\sigma^2$ is the variance of $Z$ and $R$ the correlation function. For simplicity, we consider a stationary process $Z(\mathbf{x})$, which means that correlation between $Z(\mathbf{x})$ and $Z(\mathbf{u})$ is a function of the distance between $\mathbf{x}$ and $\mathbf{u}$. Our study focuses on a particular family of correlation functions that can be written as a product of one-dimensional correlation functions $R_l$:
\[ \mbox{Cov} ( Z ( \mathbf{x} ), Z ( \mathbf{u} ) )  = \sigma^2 R ( \mathbf{x} - \mathbf{u} ) = \sigma^2 \prod_{l = 1}^d R_l ( x_l - u_l ) . \]
This form of correlation function is particularly well-suited to simplify mathematical developments in analytical uncertainty and sensitivity analyses \citep{marioo08}.
More precisely, we use the generalized exponential correlation function:
\[ R_{\mbox{\boldmath $\theta$}, \mbox{\boldmath $p$}} ( \mathbf{x} - \mathbf{u} ) = \prod_{l = 1}^d \exp ( - \theta_l |x_l - u_l |^{p_l} )  ,\]
where $\mbox{\boldmath $\theta$} = [ \theta_1, \ldots, \theta_d ]^t $ and $\mbox{\boldmath $p$} =  [ p_1, \ldots, p_d ]^t $ are the correlation parameters (also called hyperparameters) with $\theta_l \geq 0$ and $0 < p_l \leq 2$ $\;\forall\; l=1..d$.
This choice is motivated by the wide spectrum of shapes that such a function offers.

If a new point $\mathbf{x}^{\ast} = (x^{\ast}_1,\ldots,x^{\ast}_d) \in \mathcal{X}$ is considered, we obtain the following predictor and variance formulas:
 \begin{eqnarray}
 \displaystyle \mathbb{E} [  Y_{\mbox{\scriptsize Gp}}(\mathbf{x}^{*})] = f_0(\mathbf{x}^{*}) +  \mbox{\boldmath $k$}(\mathbf{x}^{*})  ^t \mbox{\boldmath $\Sigma$}_s^{-1} (Y_s - f(X_s)) \;, \label{eq_esperance} \\ 
 \displaystyle \mbox{Var}[ Y_{\mbox{\scriptsize Gp}}(\mathbf{x}^{*})] = \sigma^2  -  \mbox{\boldmath $k$}(\mathbf{x}^{*}) ^t  \mbox{\boldmath $\Sigma$}_s^{-1} \mbox{\boldmath $k$}(\mathbf{x}^{*}) \;, \label{eq_variance}
\end{eqnarray}
with $Y_{\mbox{\scriptsize Gp}}$ denoting $(Y|Y_s,X_s,\mbox{\boldmath $\beta$},\sigma,\mbox{\boldmath $\theta$}, \mbox{\boldmath $p$})$,
  \[ 
\begin{array}{lll}
\mbox{\boldmath $k$}(\mathbf{x}^{*}) & = &[\mbox{Cov}(y^{(1)},Y(\mathbf{x}^{*})), \ldots, \mbox{Cov}(y^{(n)},Y(\mathbf{x}^{*}))  ] ^t  \\
& = & \sigma^2  [  R_{\mbox{\boldmath $\theta$}, \mbox{\boldmath $p$}} (\mathbf{x}^{(1)}-\mathbf{x}^{*}), \ldots, R_{\mbox{\boldmath $\theta$}, \mbox{\boldmath $p$}} (\mathbf{x}^{(n)}-\mathbf{x}^{*}) )  ]^t  
 \end{array} 
\]
and the covariance matrix
\[ \mbox{\boldmath $\Sigma$}_s = \sigma^2 \left( R_{\mbox{\boldmath $\theta$}, \mbox{\boldmath $p$}} \left( \mathbf{x}^{(i)} - \mathbf{x}^{(j)} \right)_{i=1..n, j = 1..n}  \right) \;.\]

Regression and correlation parameters $\mbox{\boldmath $\beta$}$, $\sigma$, $\mbox{\boldmath $\theta$}$ and $\mbox{\boldmath $p$}$ are usually estimated by maximizing likelihood functions \citep{fanli06}.
This optimization problem can be badly conditioned and difficult to solve in high dimensional cases ($d>5$).
\citet{welbuc92} and \citet{marioo07} developed algorithms to build Gp metamodels on outputs that have a non-linearity depending on quite a large number of input variables.

The conditional mean (Eq. (\ref{eq_esperance})) is  used as a predictor. 
The variance formula (Eq. (\ref{eq_variance})) corresponds to the mean squared error (MSE) of this predictor and is also known as the kriging variance. 
This analytical formula for MSE gives a local indicator of the prediction accuracy. 
More generally, the Gp model provides an analytical formula for the distribution of the output variable at any arbitrary new point. 
This distribution formula can be used to develop analytical formula for uncertainty and sensitivity analyses \citep{oakoha02,oakoha04}. 
Studying several test functions and one industrial application, \citet{marioo08} showed that this analytical approach is efficient to compute the first order Sobol' indices $S_i$ (Eq. (\ref{eq:indordre1})). In addition, it provides confidence intervals for the estimates.
However, the analytical approach does not yield any direct estimation of the total Sobol' indices $S_{T_i}$ (Eq. (\ref{eq:indtotal})) and deals only with uncorrelated inputs.

\section{METHODOLOGY FOR A SPATIAL OUPUT}\label{sec:methodo}

In this section, we describe the methodology that we use to compute spatial Sobol' index maps \citep[first proposed in][]{mar08}.
We also apply this methodology to an analytical function in order to study the convergence of the algorithms. 

\subsection{General principles}

For a given $\mathbf{x}^{*}$ value  of vector $\mathbf{X}=(X_1, \ldots, X_d)$, the code output is now a deterministic function $y(\mathbf{x}^{*},\mathbf{z})$ where $\mathbf{z}$ denotes a vector of dimension $p$ of spatial coordinates. 
In this paper, we focus on two-dimensional cases. Thus, the target outputs are two-dimensional maps. Thus, $\mathbf{z}$ varies in a grid on a compact set $D_z$ of $\mathbb{R}^2$ and corresponds to an index for the outputs. 
Variables $\mathbf{X}$ and $\mathbf{z}$ are of very distinct natures: variables $X_1, \ldots, X_d$ which correspond to the inputs of the computer code, are random.
They are different for each simulation of the code. Our objective is to perform a sensitivity analysis with respect to these variables. 
Variables $\mathbf{z}$ are deterministic and  vary on a grid of size $n_z$ which corresponds to a discretization of $D_z$. 
The grid is the same for each simulation of the code and the output corresponds to the $n_z$ values $y(\mathbf{x}^{*},\mathbf{z})$ for $\mathbf{z}$ describing the grid. 
For example, the MARTHE model described in Section \ref{sec:appli} has $d=20$ input variables and yields at each simulation a map with $n_z=64\times 64=4096$ points.

Because of the different natures of variables $\mathbf{X}$ and $\mathbf{z}$, the dependency of the output with respect to these two variables is represented from two different ways. 
For a fixed value of $\mathbf{X}$, we use a projection of map $\mathbf{z} \mapsto Y(\mathbf{X},\mathbf{z})$, onto an orthonormal wavelet basis. The coefficients of the projection depend on $\mathbf{X}$. 
We select the coefficients with the largest variance and model these coefficients with respect to the $d$-dimensional input variable $\mathbf{X}$. In most applications, the dimension of $\mathbf{X}$ is quite large and each simulation of the code is time-expensive. 
Therefore, we need a method able to deal with a limited number of simulations and imput vectors of large dimension. 
In addition, the relationship between the input variables and the coefficients is expected to be highly non-linear. 
We therefore use the Gp metamodel, described in the previous section, to model the dependency of each selected coefficient with respect to $\mathbf{X}$. 

Therefore, for a given input design $X_s=\left( \mathbf{x}^{(i)} \right)_{i=1..n}$ and the $n$ corresponding simulations of the map $(y(\mathbf{x}^{(i)}, \mathbf{z}_j),j=1,\ldots, n_z)$, $i=1, \ldots, n$, the three main steps of the method are:
\begin{enumerate}
\item Decomposition of the maps $\mathbf{z} \mapsto y(\mathbf{x}^{(i)},\mathbf{z})$ onto a two-dimensional wavelet basis
\item Selection of the coefficients with the largest variance
\item Modeling of the coefficients with respect to the input variables using a Gp. 
\end{enumerate}
 At each step, we use various criteria to evaluate the performance of our procedures.
  We are then able to predict a map  $(y(\mathbf{x}^{*}, \mathbf{z}_j),j=1\ldots, n_z)$ for a new value of the input vector $\mathbf{x}^{*}$. 
  Of course, this method of map prediction (which we call a functional metamodel or also, in our case, a spatial metamodel) has the advantage, compared to the simulation of the code, to be much less time-expensive. 
  
 Finally, our functional metamodel allows us to produce maps of sensitivity analysis based upon Sobol' indices by using Monte Carlo methods (see introduction, Eqs. (\ref{eq:indordre1}) and (\ref{eq:indtotal})).
 As mentioned previously, the direct use of the computer code is impossible because of the required number of function evaluations.
 This study is restricted to the estimation of the first order indices $S_i(\mathbf{z})$ and total Sobol' indices $S_{T_i}(\mathbf{z})$ for $\mathbf{z} \in D_z$.
 These two indices allow us to quantify the individual and total influence for each input. Then, the degree of interaction with other inputs can then be deduced.
  
 \subsection{An analytical test case: The Campbell2D function}\label{sec:C2D}
  
\begin{figure}[!ht]
$$\psfig{figure=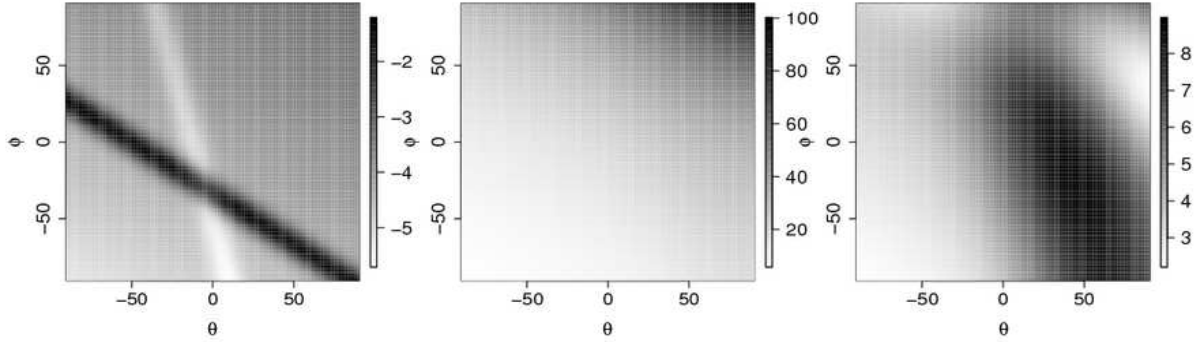,height=16.cm,width=5cm,angle=-90}$$

\caption{Three different output maps from the Campbell2D function: $\mathbf{x}=(-1,-1,-1,-1,-1,-1,-1,-1)$ (left); $\mathbf{x}=(5,5,5,5,5,5,5,5)$ (center); $\mathbf{x}=(5,3,1,-1,5,3,1,-1)$ (right).}\label{fig:campbell2D}
\end{figure}

The analytical function used in this section to perform various tests is inspired by \citet{cammck06} who considered a function with four inputs and a one-dimensional output.
It was converted to a function with eight inputs ($d=8$) and a two-dimensional output ($\mathbf{z}=(z_1,z_2)$):
\begin{equation}
	\label{eq:fct}
	\begin{array}{r}
	Y = g(\mathbf{X},z_1,z_2) = \displaystyle X_1\exp\left[-\frac{(0.8 z_1+0.2 z_2 -10X_2)^2}{60 X_1^2}\right]+(X_2+X_4)\exp\left[\frac{(0.5 z_1+0.5 z_2) X_1}{500}\right] \\
	+ \displaystyle X_5 (X_3-2) \exp\left[-\frac{(0.4 z_1+0.6 z_2 -20 X_6)^2}{40 X_5^2}\right] +(X_6+X_8) \exp\left[\frac{(0.3 z_1+0.7 z_2) X_7}{250}\right] \;,
	\end{array}
\end{equation}
where $(z_1,z_2) \in [-90,90]^2$ represent azimuthal and polar spatial coordinates and $X_i \sim {\cal U}[-1,5]$ for $i=1\ldots 8$.
This function, called the Campbell2D function, gives a spatial map as output (Figure \ref{fig:campbell2D}).
The Campbell2D function has been calibrated in order to give strong spatial heterogeneities, sometimes with sharp boundaries, and very different spatial distributions of the output values according to the $\mathbf{X}$ values.

For the Campbell2D function, it is possible to calculate the first order Sobol' indices $S_i(z_1,z_2)$.
Appendix A gives the results of these integrations.
The resulting analytical expressions (Eqs. (\ref{eq:C2DS1}) to (\ref{eq:C2DS8})) provide the exact solutions of the first order Sobol' indices.
However, analytical calculations of the total Sobol' indices $S_{T_i}(z_1,z_2)$ (Eq. (\ref{eq:indtotal})) are not possible.
We estimate $S_{T_i}(z_1,z_2)$, $i=1,\ldots,8$, by using Saltelli's Monte Carlo algorithm \citep{sal02} with $N=10^5$.
Thus, the Campbell2D function was computed $N(d+2)=10^6$ times.
The estimated errors with such large sample sizes are of the order of $5 \times 10^{-3}$ (standard deviation estimated via bootstrap).
These estimates $S_{T_i}(z_1,z_2)$ are henceforth called exact total Sobol' indices.

\begin{figure}[!ht]
     \hspace{-1cm}
\psfig{figure=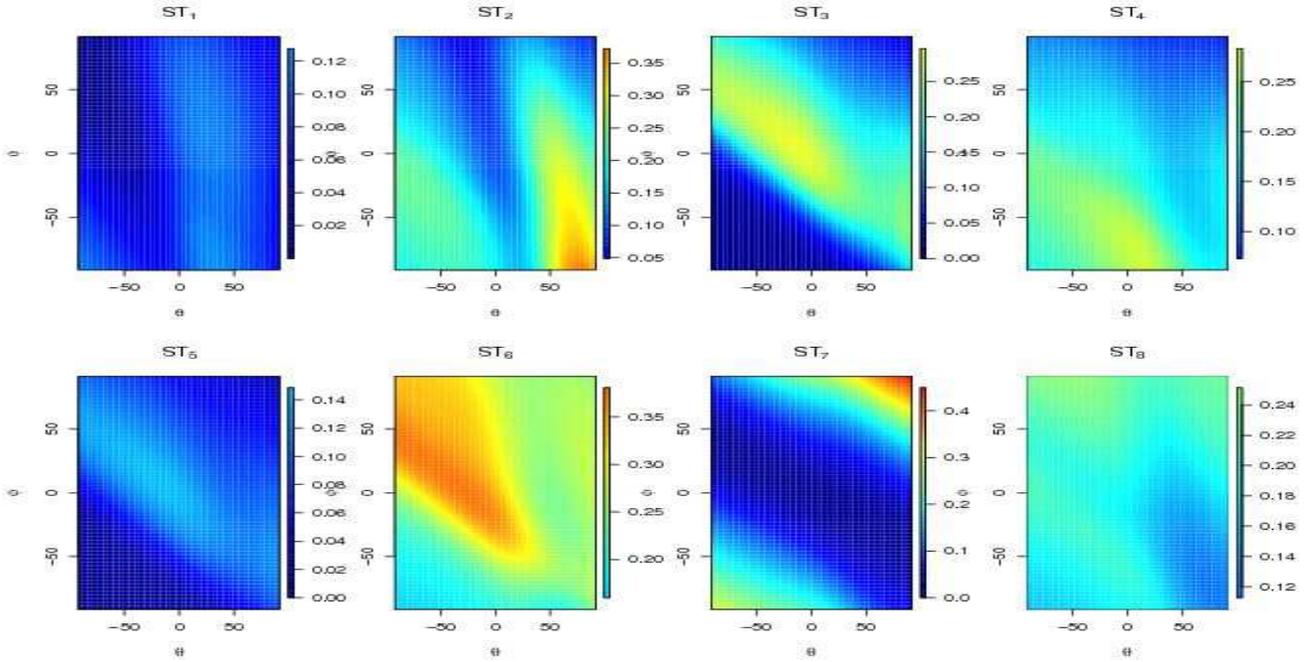,height=21.cm,width=9cm,angle=-90}
\caption{Total Sobol' indices of the $8$ input variables of the Campbell2D function, estimated by Monte Carlo algorithm. The color scales are the same for all the plots.}\label{fig:STcampbell2D}
\end{figure}

Figure \ref{fig:STcampbell2D} gives the maps of the total Sobol' index estimations.
Input $X_5$ has no influence on the output of the Campbell2D function.
Input $X_1$ has a small influence on the output of the Campbell2D function.
Input $X_3$ has a mild influence in a diagonal axis of the spatial domain.
Inputs $X_4$ and $X_8$ have mild influences in a large part of the spatial domain.
Inputs $X_2$, $X_6$ and $X_7$ have strong influences in different parts of the spatial domain (located in corners for $X_2$ and $X_7$).
Moreover, the first order Sobol' indices (maps not shown here) for $X_3$, $X_5$, $X_6$ and $X_7$ are far from the total Sobol' indices.
As shown by formula (\ref{eq:fct}), these four variables have some strong interactions (interactions between $X_3$, $X_5$ and $X_6$ and between $X_6$ and $X_7$).

\subsection{Spatial metamodeling}\label{sec:wavGp}

The spatial metamodeling process is composed of $5$ internal steps.

   \vspace{0.2cm}
   \noindent
{\bf Step 0 - Preparation of the learning sample}

   \vspace{0.2cm}
\noindent
When dealing with a large input dimension $d$, the choice of the input design $X_s=\left( \mathbf{x}^{(i)} \right)_{i=1..n}$ is very important, especially when $n$ is small.
For scalar computer model output, numerous authors stressed the strong influence of the input design on the quality of the Gp modeling \citep{koeowe96,fanli06}.
For instance, maximin Latin hypercube samples and low-discrepancy Latin hypercube samples were shown to provide good results \citep{mar08,ioobou10}.
However, building good input designs for functional output still remains an open question which could be the subject of future work.

For our tests with the Campbell2D function, we use maximin Latin hypercube samples.
Once the input design is defined, we obtain $n$ simulations of the map $\left(y(\mathbf{x}^{(i)}, \mathbf{z})\right)_{i=1..n}$ by running the numerical model.

   \vspace{0.2cm}
   \noindent
{\bf Step 1 - Spatial decomposition and selection of coefficients}

   \vspace{0.2cm}
\noindent
The spatial decomposition of the output map is made on a basis of orthogonal functions $\left\{\phi_j\right\}_{j\in \mathbb{N}^*}$:
\begin{equation}\label{eq_decompo_base}
Y (\mathbf{X},\mathbf{z}) = \mu(\mathbf{z}) + \sum^{\infty}_{j=1} {\alpha}_j(\mathbf{X}) {\phi}_j(\mathbf{z}) \text{ with } {\alpha}_j(\mathbf{X}) = \int_{D_z} [ Y(\mathbf{X},\mathbf{z}) - \mu(\mathbf{z})] {{\phi}_j(\mathbf{z})}d\mathbf{z} \;,
\end{equation}
where $\mu(\mathbf{z})=\mathbb{E}_{\mathbf{X}}[Y(\mathbf{X},\mathbf{z})]$.
We define ${Y}_K(\mathbf{X},\mathbf{z})$ as the truncated decomposition at order $K$: 
\begin{equation}\label{eq:tronc}
{Y}_K (\mathbf{X},\mathbf{z}) = \mu(\mathbf{z}) + \sum^{K}_{j=1} {\alpha}_j(\mathbf{X}) {\phi}_j(\mathbf{z}) \;.
\end{equation}
For the function basis, various wavelet bases can be considered \citep[Haar, Daubechies, Symmlet and Coiflet, see][]{mismis07} in order to optimize the compression of the local and global information.
In the following tests, we use the Daubechies basis which offered the best results.

The selection of a small number $k$ of coefficients $\alpha_j(\mathbf{X})$ to be modeled with Gp is essential.
For instance, MARTHE maps (see Section \ref{sec:appli}) and Campbell2D maps contain $n_z=64\times 64=4096$ pixels, which leads to $K=4096$ wavelet coefficients.
Modeling such a number of Gp seems intractable because the building process of one Gp is CPU time consuming \citep{marioo07}. 
It is therefore necessary to model with Gp only the most informative coefficients.
The criterion considered for selecting the coefficients involves their variance with respect to $\mathbf{X}$: 
 priority is given to coefficients which explain at most the output map variability.
 Mathematically, the new order of the coefficients $\left\{ \alpha_{1},\ldots,\alpha_{K}\right\}$ is written $\left\{ \alpha_{(1)},\ldots,\alpha_{(K)}\right\}$ following the inequalities
 \begin{equation}
 \frac{1}{n}\sum_{i=1}^n \left(\alpha_{(1)}( \mathbf{x}^{(i)})- \overline{\alpha_{(1)}}\right)^2 \geq \ldots \geq \frac{1}{n}\sum_{i=1}^n \left(\alpha_{(K)}( \mathbf{x}^{(i)}) - \overline{\alpha_{(K)}^{}}\right)^2 \;\mbox{ with }\; \overline{\alpha_j}=\frac{1}{n}\sum_{i=1}^n \alpha_j ( \mathbf{x}^{(i)})\;.
 \end{equation}
The number $k$ of Gp-modeled coefficients will be discussed in Step 3.

   \vspace{0.2cm}
   \noindent
{\bf Step 2 - Modeling the coefficients}

   \vspace{0.2cm}
\noindent
For $j=1,\ldots,K$, the model $A_j(\mathbf{X})$ used for approximating the coefficient $\alpha_j(\mathbf{X})$ is one of the following models listed below:
\begin{itemize}
\item Model 1:
the empirical mean: $A_j(\mathbf{X}) = \displaystyle\frac{1}{n}\sum_{i=1}^n \alpha_j\left(\mathbf{x}^{(i)}\right)$;
\item Model 2:
the linear regression model: 
\begin{equation}\label{eq:linear}
A_j(\mathbf{X}) = \displaystyle \beta_{0,j} + \sum_{l=1}^d \beta_{l,j}\; X_l
\end{equation}
fitted on the learning sample $\left(\mathbf{x}^{(i)},\alpha_j(\mathbf{x}^{(i)})\right)_{i=1..n}$.
We use an AIC selection process to keep only the significant terms in (\ref{eq:linear});
\item Model 3:
the Gp model of form (\ref{eq:GP}) as described in \citet{marioo07}.
The deterministic part $f_0(\mathbf{X})$ is a linear regression model as in (\ref{eq:linear}) with a selection process based on AICC (a modified AIC in order to take spatial correlations into account, see \citet{hoedav06}).
The generalized exponential function is used for the correlation function $R(\cdot)$ of the stochastic part $Z(\mathbf{X})$.
The building of this model is rather costly, especially in a  high dimensional context ($d>10$) because of the specific variable selection process proposed by \citet{marioo07}.
\end{itemize}

In the following two steps, we compare three different methodologies in order to stress the benefit of an appropriate metamodel choice:
\begin{itemize}
\item Method 1:
Model 3 for the $k$ selected coefficients and model 1 for the other coefficients
\item Method 2:
Model 2 for the $k$ selected coefficients and model 1 for the other coefficients
\item Method 3:
Model 3 for the $k$ selected coefficients, model 2 for the $k'$ following coefficients ($k' \gg k$) and model 1 for the $K-k-k'$ other coefficients.
For the Campbell2D function, setting $k'$ to $500$ is a heuristic choice based upon the observation that, in the case studied, the information in terms of variability is explained by $10\%$ of coefficients.
More generally, a convergence study can be made in order to find a suitable value for $k'$.
\end{itemize}
We now define $\widehat{Y}_{K,k} (\mathbf{X},\mathbf{z})$ the approximation of ${Y}_K (\mathbf{X},\mathbf{z})$ (Eq. (\ref{eq:tronc})) using one of the three previous methods.

Several adequacy criteria can be used to measure the discrepancy between the function $Y (\mathbf{X},\mathbf{z})$ and its approximation $\widehat{Y}_{K,k} (\mathbf{X},\mathbf{z})$.
We use the mean absolute error, the maximal error and the mean squared error but restrict our presentation to mean squared error results for the sake of consistency.
The mean squared error $\mbox{MSE}(\mathbf{X})$ is written
\begin{equation}\label{eq:MSE}
\mbox{MSE}(\mathbf{X}) = \int_{D_z} \left[ Y(\mathbf{X},\mathbf{z}) - \widehat{Y}_{K,k} (\mathbf{X},\mathbf{z}) \right]^2 d\mathbf{z} \;.
\end{equation}
$\mbox{MSE}(\mathbf{X})$ is estimated by integrating over the $n_z$ grid.
For a fixed value of $\mathbf{X}$, this criterion measures the restitution quality in the mean of the overall map. We denote by MSE the expectation (with respect to the variable $\mathbf{X}$) of $\mbox{MSE}(\mathbf{X})$. When it is possible, we provide new simulations of the map $ Y(\mathbf{X},\mathbf{z})$ for randomized values of $\mathbf{X}$, and we use this test sample to estimate the MSE. For some applications, this is not possible and cross-validation methods can be used to estimate the MSE (see Section \ref{sec:appli}). 

The MSE can also be obtained by first integrating $\left[ Y(\mathbf{X},\mathbf{z}) - \widehat{Y}_{K,k} (\mathbf{X},\mathbf{z}) \right]^2$ over $\mathbf{X}$ and then by taking the expectation with respect to $\mathbf{z}$.
From the MSE, we also define the predictivity coefficient $Q_2$ which gives us the percentage of the mean explained variance of the output map:
\begin{equation}\label{eq:Q2}
Q_2 = 1 - \frac{\mbox{MSE}}{\mathbb{E}_{\mathbf{z}}\left\{\mbox{Var}_{\mathbf{X}}\left[ Y(\mathbf{X},\mathbf{z}) \right]\right\}} \;.
\end{equation}
The variance is taken with respect to $\mathbf{X}$ because we are interested in the variability induced by the model input vector $\mathbf{X}$. 
$Q_2$ corresponds to the coefficient of determination $R^2$ computed in prediction (on a test sample or by cross-validation).

   \vspace{0.2cm}
\noindent
{\bf Step 3 - Choosing $k^{*}$, an optimal value for $k$}

   \vspace{0.2cm}
   \noindent
We perform simulations using the Campbell2D function and study convergence of MSE (Eq. (\ref{eq:MSE})) as function of $k$.
Our goal is to compare the three methods proposed in step 2, then to heuristically find an optimal value $k^{*}$ for $k$.
Indeed, there is a trade-off between keeping $k$ small and minimizing the MSE.
The MSE is computed using a test sample of $1000$ independent Monte Carlo simulations, giving $1000$ output maps.

\begin{figure}[!ht]
$$\psfig{figure=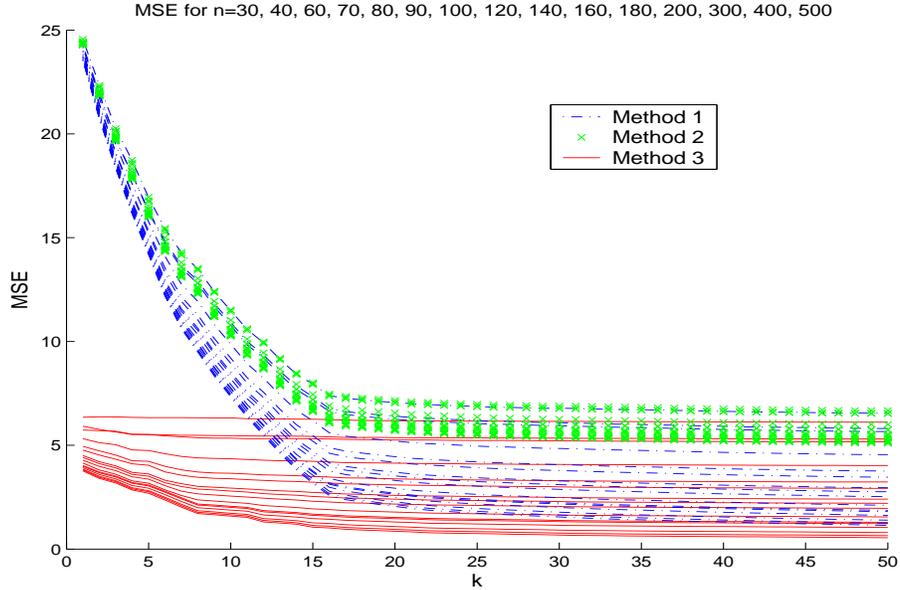,height=8.cm,width=12cm}$$
\caption{For the Campbell2D function, MSE convergence (as function of $k$) for the three methods and for various learning sample sizes ($n=30, 40, 50, 60, 70, 80, 90, 100, 120, 140, 160, 180, 200, 300, 400, 500$).}\label{fig:MSE_k}
\end{figure}

Figure \ref{fig:MSE_k} gives the MSE results as function of $k$ for different values of the learning sample size $n$.
For each method, the MSE curves regularly turn downward as $n$ increases.
As expected, method 3, which is the richest in terms of model complexity, gives the best results, especially for small values of $k$.
The usefulness of Gp is proved as we see that method 2 performs badly.
It is certainly caused by the behavior of the first selected coefficients, which offer strong and non-linear variations: linear models are irrelevant for modeling these coefficients.
For each method, the convergence is reached for $k$ around $20$ - $25$.
We decided to fix the optimal value at $k^{*}=30$, which is a reasonable number of Gp models to be built.

In real applications, this methodology for choosing $k^{*}$ can be applied even if the learning sample size $n$ is limited.
For a fixed $n$, we look for a stabilization of the MSE.
If this convergence is not reached, we use a predefined maximal value for $k$.

It should be noted that if new model runs are available, the analyst has to repeat the process to choose $k^{*}$.
However, in order to gain some analysis time, we can leave unchanged the ordering of coefficients, which has been obtained with the first set of simulations.
In addition, we can just update the predictor (Eq. (\ref{eq_esperance})) by keeping the initial estimation of the correlation parameters (which is the most cpu time consuming step).
Such choices have to be made with care.

   \vspace{0.2cm}
\noindent
{\bf Step 4 - Convergence as function of the learning sample size $n$}

\begin{figure}[!ht]
$$\psfig{figure=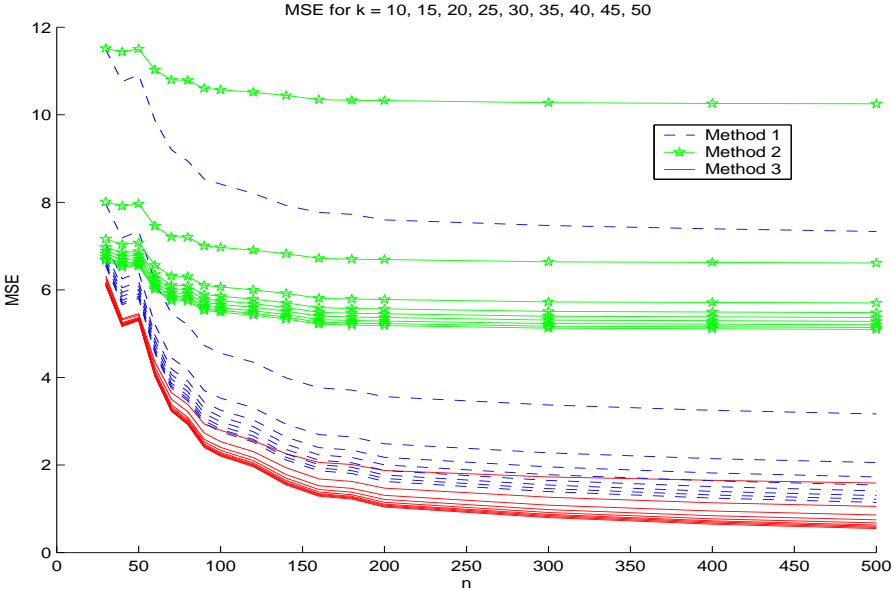,height=8.cm,width=12cm}$$
\caption{For the Campbell2D function, MSE convergence (as function of $n$) for the three methods and for various numbers $k$ of Gp-modeled coefficients.}\label{fig:MSE_n}
\end{figure}

 \vspace{0.2cm}
 \noindent
 Finally, it is important to study the convergence of the adequacy criteria as function of the learning sample size $n$.
 It would allow us to eventually prescribe the need to make new simulations with the code.
 For the Campbell2D function, Figure \ref{fig:MSE_n} gives the MSE results as function of $n$ for different values of $k$.
For each method, the MSE curves regularly turn downward as $k$ increases.
 In real applications, one can restrict this to the visualization of the $k^{*}$ curves.
 
 Method 2 performs badly and the stabilization of its curves is obtained earlier.
 Indeed, adding simulations does not improve the linear models fitted on the $k$ coefficients.
 For methods 1 and 3, the curve stabilization is not reached at $n=500$.
 MSE would decrease for larger values of $n$, but this decrease becomes slower from $n=200$ and MSE results are rather satisfactory for this value $n=200$.
 In terms of predictivity coefficient (Eq. (\ref{eq:Q2})), we obtain $Q_2=96.6 \%$ for $n=500$ and $Q_2=92.9 \%$ for $n=200$.
 For methods 1 and 3, increasing $k$ and $n$ leads to  a systematic decrease of the MSE.
 It can therefore be argued that MSE tends to zero and that our methodology converges. 
 
 In real applications, if no additional simulation can be made, this step can be optional.
However, in the opposite case, these curves would help us to decide if our simulation number is sufficient and which method we have to choose.
 Moreover, knowing that method 3 can be costly, we can decide to choose method 1 if their MSEs are similar.
 In practical terms, we start from an initial $n_0$ (random selection of $n_0$ simulations among the $n$ simulations) and randomly add simulations until $n$.
The choice of a low-discrepancy sequence would also allow the space-filling properties of the design to be kept while increasing $n$.
 
In conclusion, by analyzing all these convergence plots, we choose in the next section to use a learning sample size $n=200$ and to model $k^{*}=30$ Gps using method 3 in order to compute Sobol' indices.



   \vspace{0.2cm}
   \noindent
{\bf Coarse estimation of the computational time of the different steps}

 \vspace{0.2cm}
 \noindent
 Concerning the computational time needed to carry out all our methodology, the most costly steps are the construction of each Gp metamodel for the $k^{*}$ wavelet coefficients and the validation step (i.e. computation of MSE or $Q_2$ by cross validation). All the other steps such as the wavelet decomposition, the selection of coefficients or the prediction of the functional metamodel for any new input value are negligible in terms of computational time.
 
So, the first main difficulty is the hyperparameter estimation of the $k^{*}$ Gp metamodels. Indeed, each computation of the likelihood requires the inversion of correlation matrix and consequently, the maximum likelihood estimation can be CPU time consuming. In the case of $10$ inputs for example and a few hundreds of simulations, a Gp modeling usually requires several minutes on a standard PC (Pentium 4, $1.8$GHz). So, for tens of coefficients to be modeled, the step 2 can take one hour.

The second difficulty is the validation step, also because of the time required by the maximum likelihood estimation. To reduce its computational cost, a k-fold cross-validation is preferable in practice and limits the time required for cross validation to just a few hours. Another solution is to leave unchanged the hyperparameters of Gp at each loop of cross validation. Only the Gp predictor is updated. The cross validation is then a little biased but, for a few hundreds of simulations, this bias becomes quickly negligible.

As a conclusion, only the step of the Gp modeling is computationally expensive. For instance, in the Campbell2D function study, with $d=8$ inputs, $n_z=4096$ pixels, $k$ ranging from $10$ to $50$ Gp models and $n=200$ simulations, the metamodeling process from steps 1 to 3 (without the convergence plot in function of $n$) required approximately one day. For the MARTHE test case, with $d=20$, $n_z=4096$, $n=300$, $k=100$ and with a $10$-fold cross-validation process, the computation of all the Sobol' indices has required approximately two days.
These operational cost may appear to be high but this process is only made once to obtain a full functional metamodel.
     Afterwards, any evaluation of the metamodel
    will require a negligible computational time compared to a simulation of the initial MARTHE simulator.

\subsection{Global sensitivity analysis}\label{sec:GSAC2D}

At this stage, we have a functional metamodel allowing us to predict new output concentration maps for any new set of input variables.
This metamodel has been obtained with only $n=200$ computations with the Campbell2D function.
To estimate Sobol' indices of the overall output map of the Campbell2D function, we then perform thousands of simulations on our functional metamodel.
This method is called hereafter the functional metamodel-based approach.

Note that there is no direct link between the Sobol' indices for the wavelet coefficients and the Sobol' indices for the output. Indeed from (\ref{eq:tronc}), we have
$$\displaystyle Y_K(\mathbf{X},\mathbf{z})=\mu(z)+\sum_{j=1}^K \alpha_j(\mathbf{X}) \phi_j(\mathbf{z}),$$
where $(\alpha_j(\mathbf{X}))_{1\leq j\leq K}$ denote the wavelet coefficients. The sensitivity map with respect to the variable $X_i$ is $\displaystyle S_i(\mathbf{z})= \frac{\mbox{Var}\{\mathbb{E}[Y_K(\mathbf{X},\mathbf{z})] |X_i\}}{\mbox{Var}[Y_K(\mathbf{X},\mathbf{z})]}$. Hence,
$$S_i(\mathbf{z})=  \frac{\sum_{j,l=1}^K \mbox{Cov}\{\mathbb{E}[\alpha_j(\mathbf{X}) |X_i],\mathbb{E}[\alpha_l(\mathbf{X}) |X_i]\} \phi_j(\mathbf{z})\phi_l(\mathbf{z})}{\mbox{Var}[Y_K(\mathbf{X},\mathbf{z})]}.$$

If the functions $( \phi_j(\mathbf{z}))_{1\leq j\leq K}$ have disjoint supports,  all the terms with $l\neq j$ in the above formula equal zero and Sobol' indices for the wavelet coefficients could be used to compute Sobol' indices for the output. 
In this paper, this is not the case as we use the Daubechies basis for these functions.
This basis gave much better results than bases with disjoint supports functions (such as the Haar basis).

Thus, to estimate Sobol' indices of the overall output map of the Campbell2D function, we perform thousands of simulations on our functional metamodel. Because of constraints of memory allocation (due to the size of the output map and our vectorial programming constraints), it is not possible to use Saltelli's Monte Carlo algorithm \citep{sal02}.
Therefore, we use the following procedure for each of the $n_z$ nodes of the grid:
\begin{itemize}
\item For the variance of the conditional expectation of each input variable $X_i$ ($i=1,\ldots,8$), we perform $1000$ Monte Carlo computations to estimate $\mathbb{E}(Y|X_i)$ (integration over $7$ dimensions) and $200$ Monte Carlo computations to estimate $\mbox{Var}[\mathbb{E}(Y|X_i)]$ (integration over one dimension).
\item For the variance of the conditional expectation of each  $X_{\sim i}$ ($i=1,\ldots,8$), we perform $100$ Monte Carlo computations to estimate $\mathbb{E}(Y|X_{\sim i})$ (integration over one dimension) and $1000$ Monte Carlo computations to estimate $\mbox{Var}[\mathbb{E}(Y|X_{\sim i})]$ (integration over $7$ dimensions).
\item The variance of the output $\mbox{Var}(Y)$ is obtained using $2\times 10^4$ simulations (integration over $8$ dimensions).
\item Thus, the first order Sobol' index estimates (noticed $S_i^{\mbox{\tiny Gp}}$) are obtained from Eq. (\ref{eq:indordre1}) and the total Sobol' index estimates (noticed $S_{T_i}^{\mbox{\tiny Gp}}$) are obtained from Eq. (\ref{eq:indtotal}).
\end{itemize}
Finally, we obtain the Sobol' indices $S_i^{\mbox{\tiny Gp}}(\mathbf{z})$ and $S_{T_i}^{\mbox{\tiny Gp}}(\mathbf{z})$ for all the $n_z$ grid points.

\begin{figure}[!ht]
$$\psfig{figure=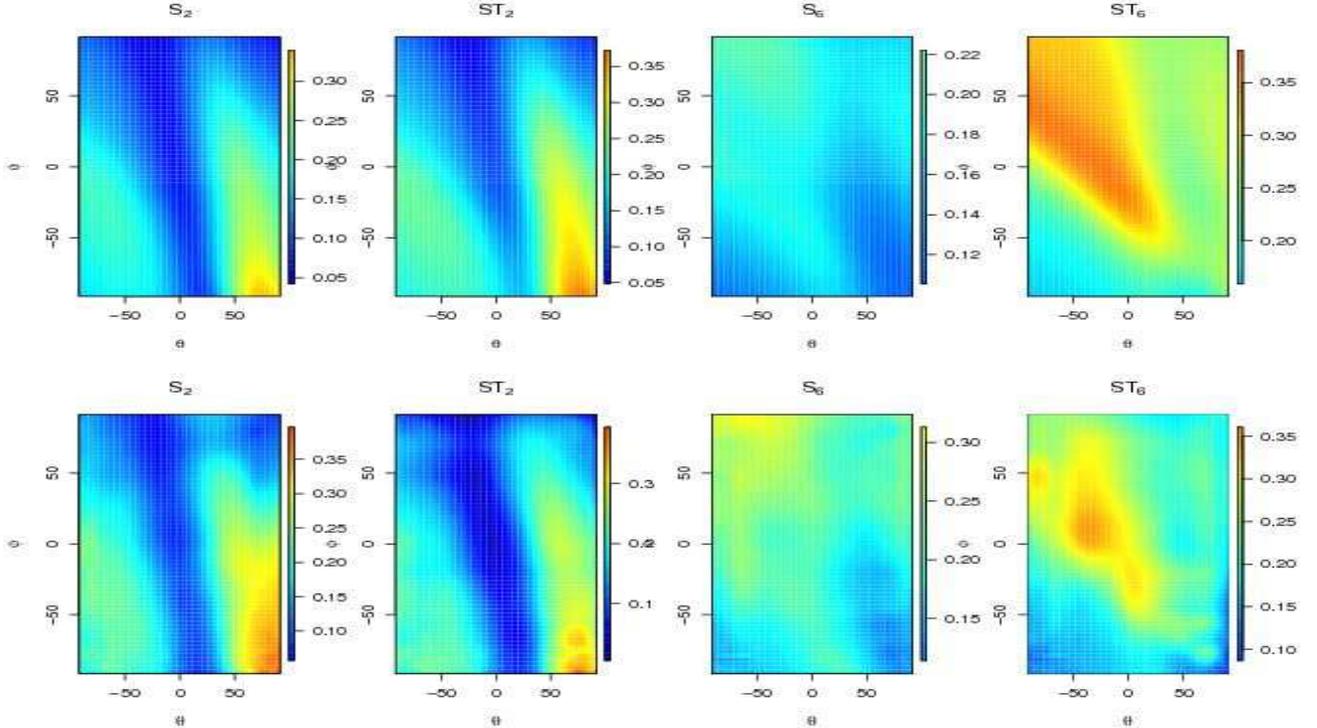,height=21cm,width=10cm,angle=-90}$$
\caption{For the Campbell2D function and variables $X_2$ and $X_6$, comparison between exact first order and total Sobol' indices (top) and functional metamodel-based Sobol' indices (bottom). The color scales are the same for all the plots.}\label{fig:C2Dsobol}
\end{figure}

Figure \ref{fig:C2Dsobol} shows the Sobol' index maps for $X_2$ and $X_6$, which are the most influential input variables in the Campbell2D function (see Fig. \ref{fig:STcampbell2D}).
Results for $X_2$ are particularly convincing: first order and total sensitivity values obtained with the functional metamodel-based approach are accurate everywhere in the spatial domain $D_z$.
Results for $X_6$ are fairly good for the first order Sobol' index and less precise for the total Sobol' index.
However, the spatial influence zone of $X_6$ in the upper left corner is well retrieved by the functional metamodel-based approach. 
In fact, $X_2$ corresponds to a solely influential input variable while $X_6$ has significant interactions with other input variables (mainly with $X_3$).
Therefore, because of a more difficult Gp fitting process, the Gp models of the wavelet coefficients of $X_6$ are less precise than the Gp models of the wavelet coefficients of $X_2$.
However, we argue that the important information is present in the spatial Sobol' map of $S_{T_6}^{\mbox{\tiny Gp}}(\mathbf{z})$.

For all the input variables, the relative mean absolute errors of the first order Sobol' indices,
\begin{equation}\label{eq:rMAE}
\mbox{rMAE}(S_i)=\frac{\mathbb{E}_{\mathbf{z}}|S_i^{\mbox{\tiny Gp}}(\mathbf{z})-S_i(\mathbf{z})|}{\mathbb{E}_{\mathbf{z}}[S_i(\mathbf{z})]} \;,
\end{equation}
were estimated for $i=1,\ldots,8$ (see Table \ref{tab:errSi}).
The results of Table \ref{tab:errSi} show that the estimations of the sensitivity maps for $X_2$ and $X_6$ correspond to one of the most difficult cases. Figure \ref{fig:C2Dsobol} shows that a mean absolute error of a $15\%$-order is quite satisfactory in terms of sensitivity maps.
Therefore, all the results for the other input variables show that our functional metamodel-based approach gives precise results.
Note that the rMAE value for $X_5$ is not given because $S_5(z_1,z_2)=0 \;\; \forall (z_1,z_2) \in [-90,90]^2$, and the denominator in Eq (\ref{eq:rMAE}) is equal to zero.

\begin{table}
  \centering
\caption{For the Campbell2D function, relative mean absolute errors (in percent) of the first order sensitivity indices estimated via the functional metamodel-based approach.}\label{tab:errSi}
\begin{tabular}{cccccccc}
&&&&&&&\\
\hline
$X_1$ & $X_2$ & $X_3$ & $X_4$ & $X_5$ & $X_6$ & $X_7$ & $X_8$ \\
\hline
$8.75$ & $16.25$ & $16.35$ & $12.8$ & --- & $13.17$ & $11.80$ & $9.96$ \\
\hline
\end{tabular}
\end{table}

In conclusion, we have shown the efficiency of this new spatial global sensitivity analysis method for this analytical and relatively complex test function: all sensitivity index spatial maps have been obtained using only $n=200$ computations of the Campbell2D function.

\section{APPLICATION}\label{sec:appli}

\subsection{The environmental problem}\label{sec:kurchatov}

In the period between 1943 and 1974 radioactive waste was buried in eleven temporary repositories built on a specially allocated site at the RRC Kurchatov Institute (KI) in the Moscow area (Russia). 
The site used for radioactive waste interim storage covers an area of about 2 hectares and is situated near the KI external perimeter in the immediate vicinity of the city's residential area. A radioactive survey of the site and its adjacent area performed in the late 1980s - early 1990s and in 2002 showed that radioactive contamination is not only present on the surface but has a tendency to spread into the groundwater. 
The porous media of the site is represented principally by sands alternatively with clays that form several horizontal superposed aquifers. 
To analyze radioactive contamination of groundwater, about a hundred exploration wells were drilled on the site. As a result of the survey, it was discovered that contamination of groundwater concerns mainly connected to $^{90}$Sr. 
Since the radiation survey results have demonstrated the necessity to clean up the site, rehabilitation activities on radwaste removal and liquidation of old repositories were performed at the site between 2002 and 2006. A network of observation wells is used to control groundwater conditions of the two upper aquifers. This network consists of twenty observation wells for the upper moraine aquifer and nine for the second Jurassic aquifer.
It is used for a regular recording of groundwater levels, its chemical and radionuclide composition \citep[see][]{velpon07}.

\begin{figure}[!ht]
$$\hspace{-1cm}
\psfig{figure=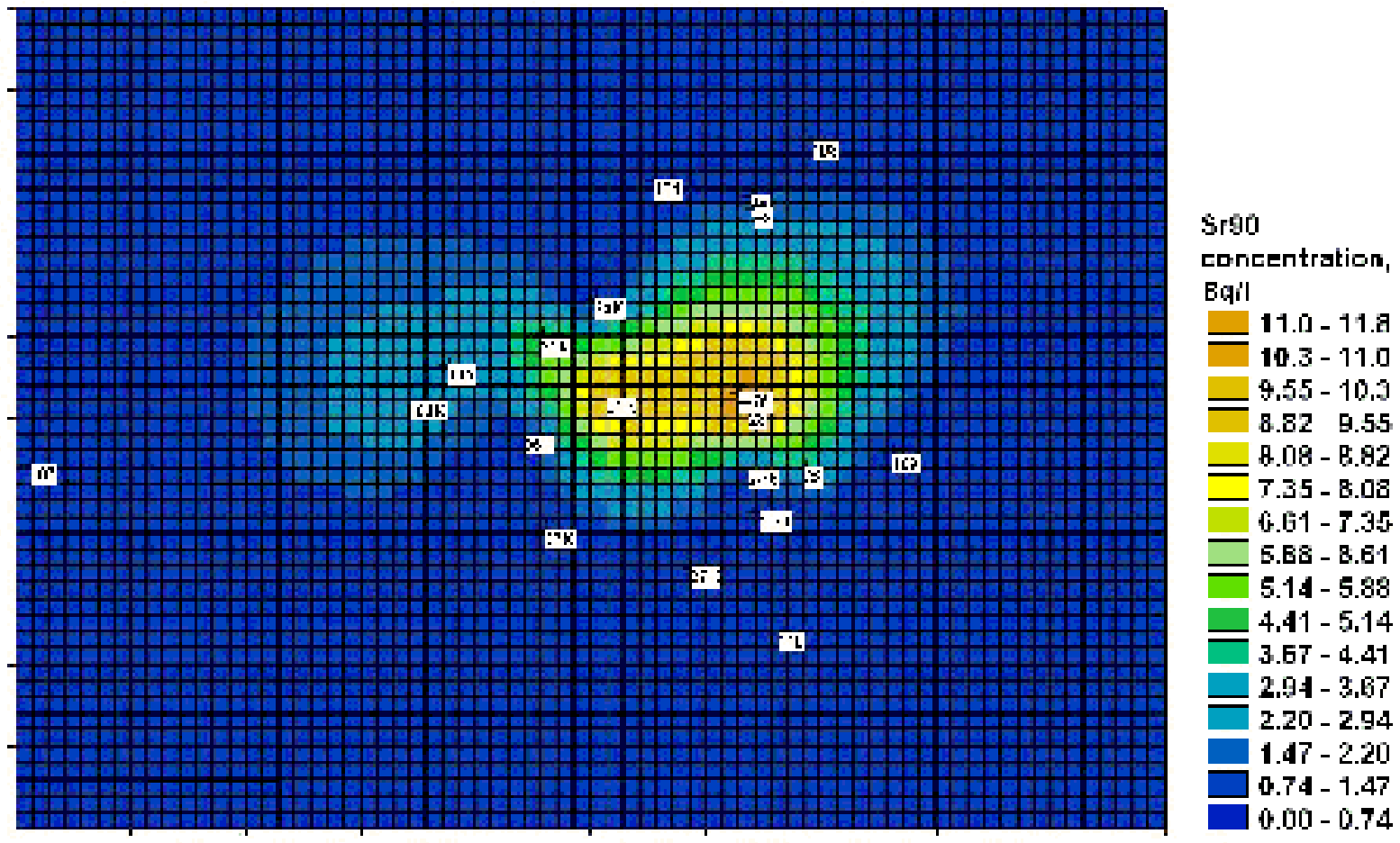,height=6.cm,width=9cm}
\psfig{figure=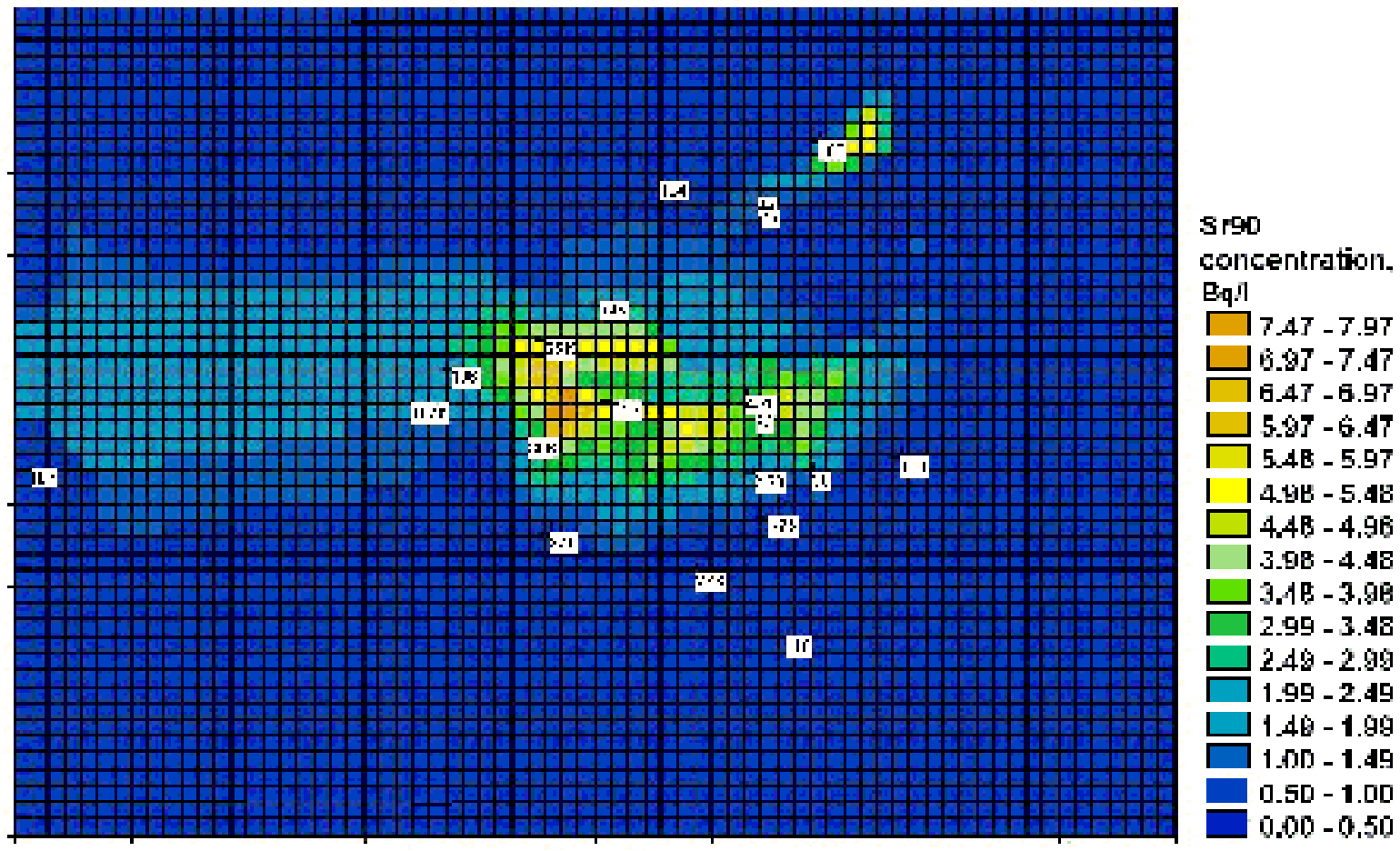,height=6.cm,width=9cm}$$

\caption{Initial (left, 2002) and predicted (right, 2010) $^{90}$Sr concentrations (hot colors represent higher levels of concentration). Initial concentrations range from $0$ to $12$ Bq/l while final concentrations range from $0$ to $8$ Bq/l. The small white rectangles represent the location of the observation wells.}\label{fig:conc}
\end{figure}

A numerical model of $^{90}$Sr transport in groundwater was developed for the RRC Kurchatov Institute (KI) radwaste disposal site \citep{volioo08}. It aimed to provide a correct prediction of further contamination plume spreading since 2002 (using an interpolated concentration map) and up to the end of the year 2010, to show the risks associated with contamination and to serve as a basis for engineering decision-making.
The numerical model was constructed using the MARTHE hydrogeological program package (developed by BRGM, the French Geological Survey). It is a three-dimensional combined transient flow and transport convection-dispersion model taking into account sorption and radioactive decay. 
Three layers were singled out; horizontal, vertical and temporal meshes were chosen in accordance with the migration characteristics of the sand.
Initial concentration plume in 2002 and spreading prediction made for the year 2010 are shown in Figure \ref{fig:conc}.
As can be seen, contamination plume predicted for the year 2010 is not uniform and is more diffused than the initial one. This is due, above all, to the influence of intensive infiltration assigned to several zones of the model domain that results in local dispersion of the contamination plume.

It has been shown in \citet{volioo08} that the shape of the predicted contamination plume depends on the model input values (hydraulic conductivity, infiltration parameters, sorption distribution coefficients, etc.).
Indeed, a large part of the model input variables are exposed to some uncertainty, since their values have been obtained through expert judgment, model calibration, field experiments and laboratory experiments. 
These uncertainties lead to uncertainties in model prediction. 
In order to evaluate the degree of input influence on the resulting contamination plume shape and concentration values predicted in observation wells, it was proposed to perform global sensitivity analysis on this numerical model (called MARTHE in the following sections).

\subsection{Global sensitivity analysis on scalar outputs}

From expert judgment and laboratory experiments, probability distributions (uniform and Weibull laws) were assigned to $20$ random input variables of MARTHE.
$300$ Monte Carlo simulations, based upon Latin hypercube sampling of the input variables \citep{mckbec79}, were performed (requiring four calculation days).
For each simulated set of input variables, MARTHE computes transport equations of $^{90}$Sr and predicts the evolution of $^{90}$Sr concentration. 
The $20$ uncertain model parameters are the permeability of different geological layers composing the simulated field, longitudinal and  transverse dispersivity coefficients, and sorption distribution coefficients.
To perform global sensitivity analysis and in particular to compute Sobol' indices, previous studies have concentrated on $20$ scalar outputs of $^{90}$Sr concentration values, predicted for the year 2010, in $20$ piezometers located on the waste repository site.

Because of the long computing time of MARTHE and of the non-linearity of the relationships between inputs and outputs, \citet{volioo08} proposed to fit a metamodel (based upon the boosting of regression trees) on each output using the learning sample ($300$ observations).
The boosting trees method consists of a sequential construction of weak models (here regression trees with low interaction depth), that are then aggregated. 
This leads to a relatively efficient metamodel (but difficult to interpret).
Then Sobol' indices were computed by intensive Monte Carlo simulations using this metamodel.
In \citet{marioo07}, each output was modeled by a Gp metamodel. The Gp metamodel outperforms the linear regression and the boosting regression trees metamodel in terms of predictivity of the output values.

As a result of these sensitivity analyses, we note that the calculated concentrations at the piezometric locations are mainly influenced by the distribution coefficient of $^{90}$Sr in the first and second layers of the domain and by the intensity infiltration in the pipe leakage zones, and to a lesser extent by the hydrodynamic parameters (dispersivity, porosity, etc.).
However, we are aware that spatial information has been lost in these analyses, due to the limited amount of output values that we have considered (concentrations located at $20$ locations).
Our goal was then to compute Sobol' indices in the whole spatial concentration map, predicted by the model for 2010.

\subsection{Global sensitivity analysis on the output concentration map}

The methodology presented in the previous section was then applied to MARTHE. 
Remember that this model contains $d=20$ input random variables and that the $n=300$ simulations have been performed following a Latin hypercube sample in a previous work.
In previous studies, $20$ scalar output variables had been considered and we hoped to obtain more information by using all the spatial information contained in the maps.
We used the $300$ spatial output maps, discretized in $n_z=4096$ pixels and predicting the $^{90}$Sr concentration values in 2010.

\begin{figure}[!ht]
$$\hspace{-0.5cm}
\psfig{figure=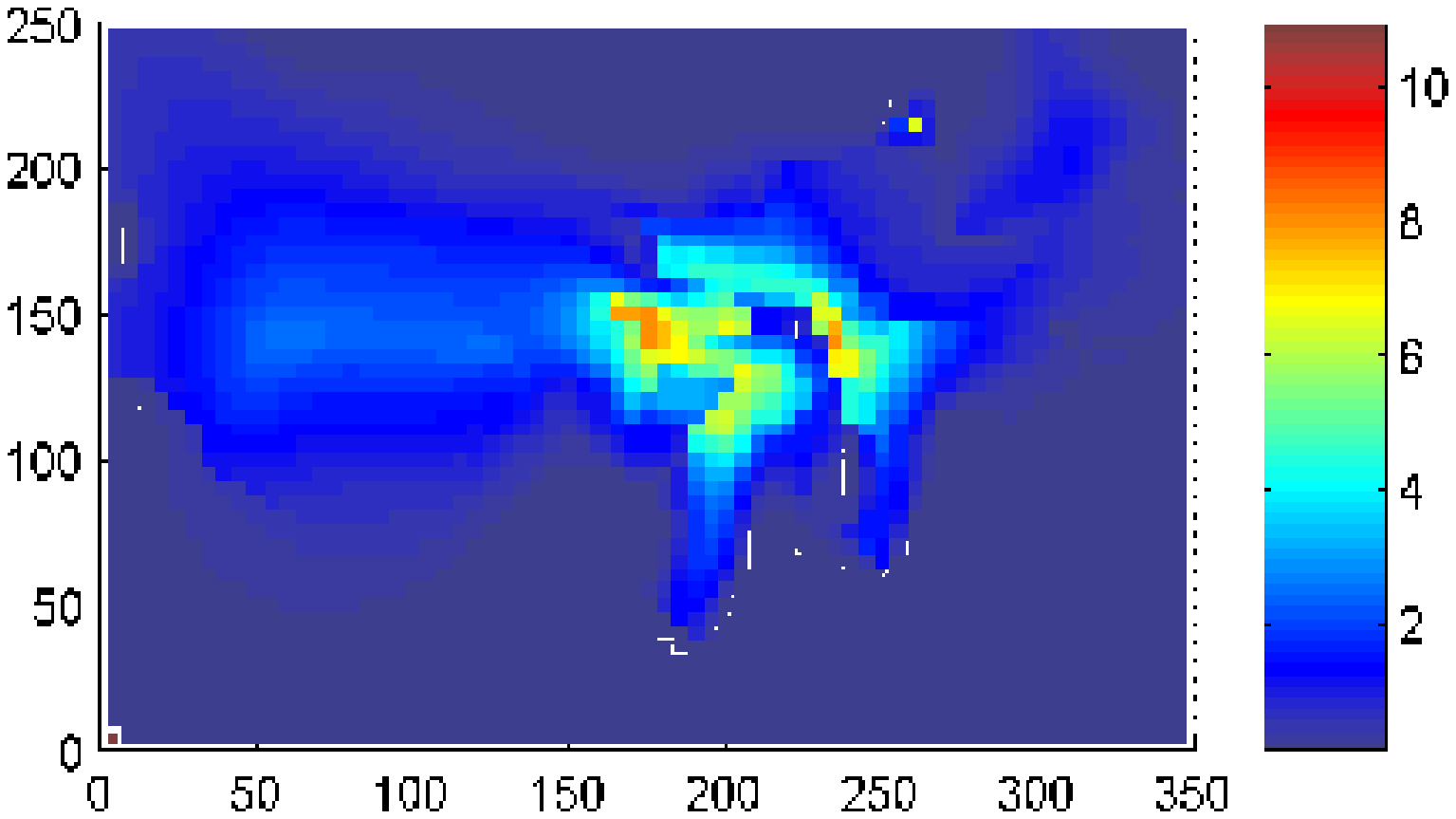,height=4.cm,width=6cm}
\psfig{figure=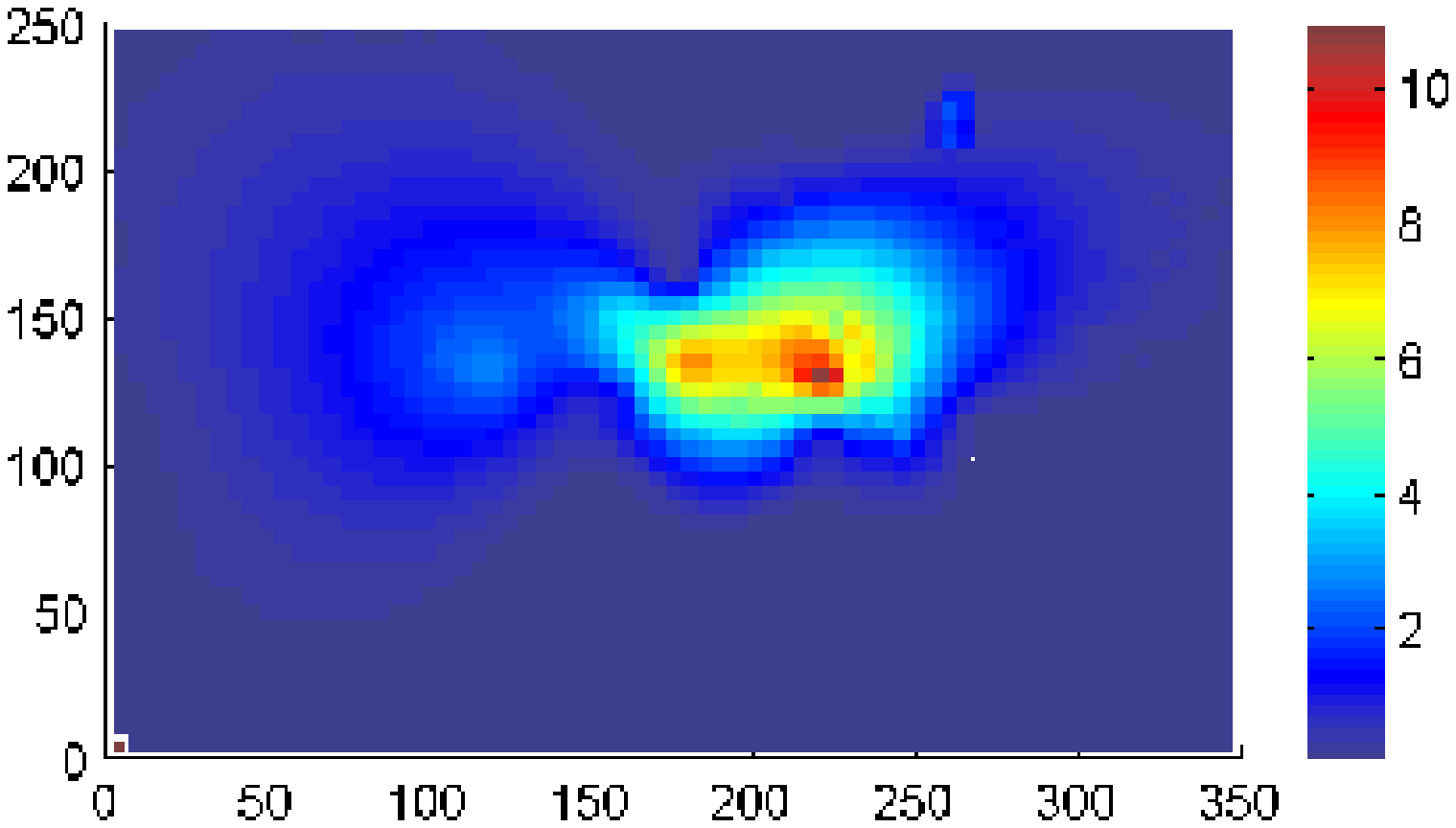,height=4.cm,width=6cm}
\psfig{figure=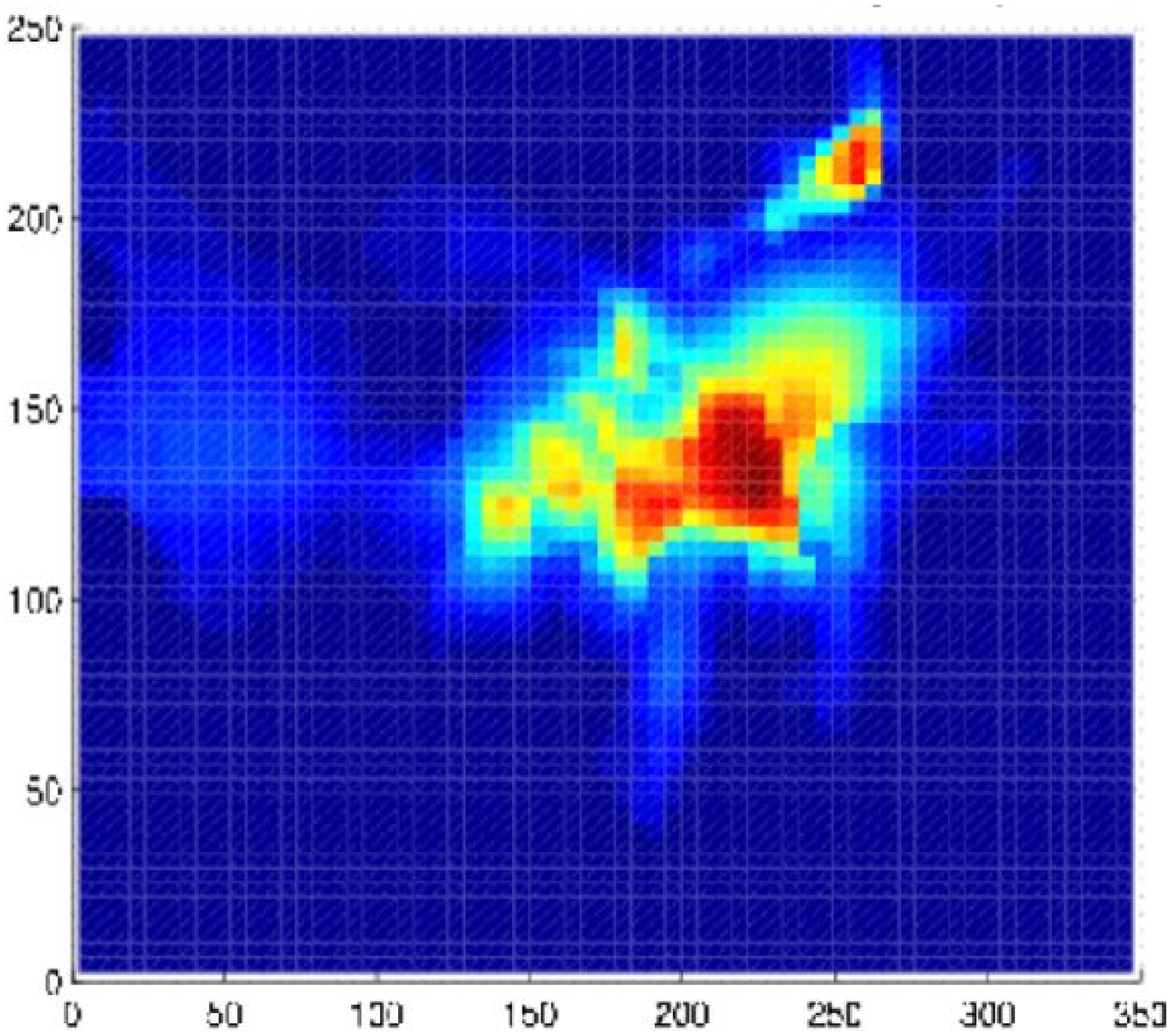,height=4.cm,width=5.5cm}$$
\hspace{2cm} (a) \hspace{5.cm} (b) \hspace{5.5cm} (c)
\caption{(a) and (b): Two final concentration maps of MARTHE (units in Bq/l). (c): Variance of the $300$ concentration maps (colors are in logarithmic scales, ranging from $0$ to $10$).}\label{fig:cartes}
\end{figure}

Figure \ref{fig:cartes} (a) and (b) shows two output maps and exemplifies the potential variability between the maps and their contour irregularity.
Another output map (Figure \ref{fig:conc}, right) confirms this observation.
The variance of the $300$ maps (Figure \ref{fig:cartes} (c)) allows us to illuminate the strong-variability zones (central spot), the mild-variability zones (on the left and at the top of the central spot) and the zones with no variability where the concentration values are equal to zero (the major part of the maps).
All this corroborates the need for a non-trivial functional metamodel, such as our wavelet-Gp based metamodel decribed in Section \ref{sec:wavGp}.

As step 0 was already done, we applied the remaining steps of the spatial global sensitivity analysis methodology (see Section \ref{sec:methodo}) using our learning sample of size $n=300$.
From steps 2 and 3, we retained method 3 with the choice of $k^{*}=100$ modeled coefficients with Gp: the stabilization of MSE was observed for this value of $k^{*}$.
The number of coefficients modeled with linear models is $k'=900$.
Step 4 was not applied to this application case.
Indeed, MARTHE simulations have been performed in a previous study \citep{volioo08} and the computer code is no longer available.
Therefore, no additional point could be added and step 4 would be useless.

In the MARTHE application, no test basis was available to compute the MSE in prediction.
The MSE estimate was obtained via a $10$-fold cross-validation technique.
The learning sample was randomly divided into $10$ sub-samples.
Then, we iterated $10$ times the following process: learning the functional metamodel on $9$ sub-samples and estimating the MSE on the remaining sub-sample.
Our final MSE estimate is the mean of the $10$ obtained MSE values: MSE$=0.039$.
In terms of predictivity coefficient (Eq. (\ref{eq:Q2})), we obtain $Q_2=72.1 \%$.
All the details of this study are given in \citet{mar08}.

At present, the functional metamodel can be used to estimate first order and total Sobol' indices.
We used Saltelli's Monte Carlo algorithm (as for the total Sobol' indices in Section \ref{sec:C2D}) with $N=10^3$.
Indeed, the low computational cost of our metamodel makes it possible to carry thousands of simulations, but not billions because of memory allocation problems (see Section \ref{sec:GSAC2D}).
The final computation cost of Saltelli's algorithm is $N (d+2)$, which leads to a number of $22000$ metamodel-based simulations in our case.
As a final result, we obtain $20$ maps of first order Sobol' indices and $20$ maps of total Sobol' indices (two maps for each input).

\begin{figure}[!ht]
     \begin{center}
$$\psfig{figure=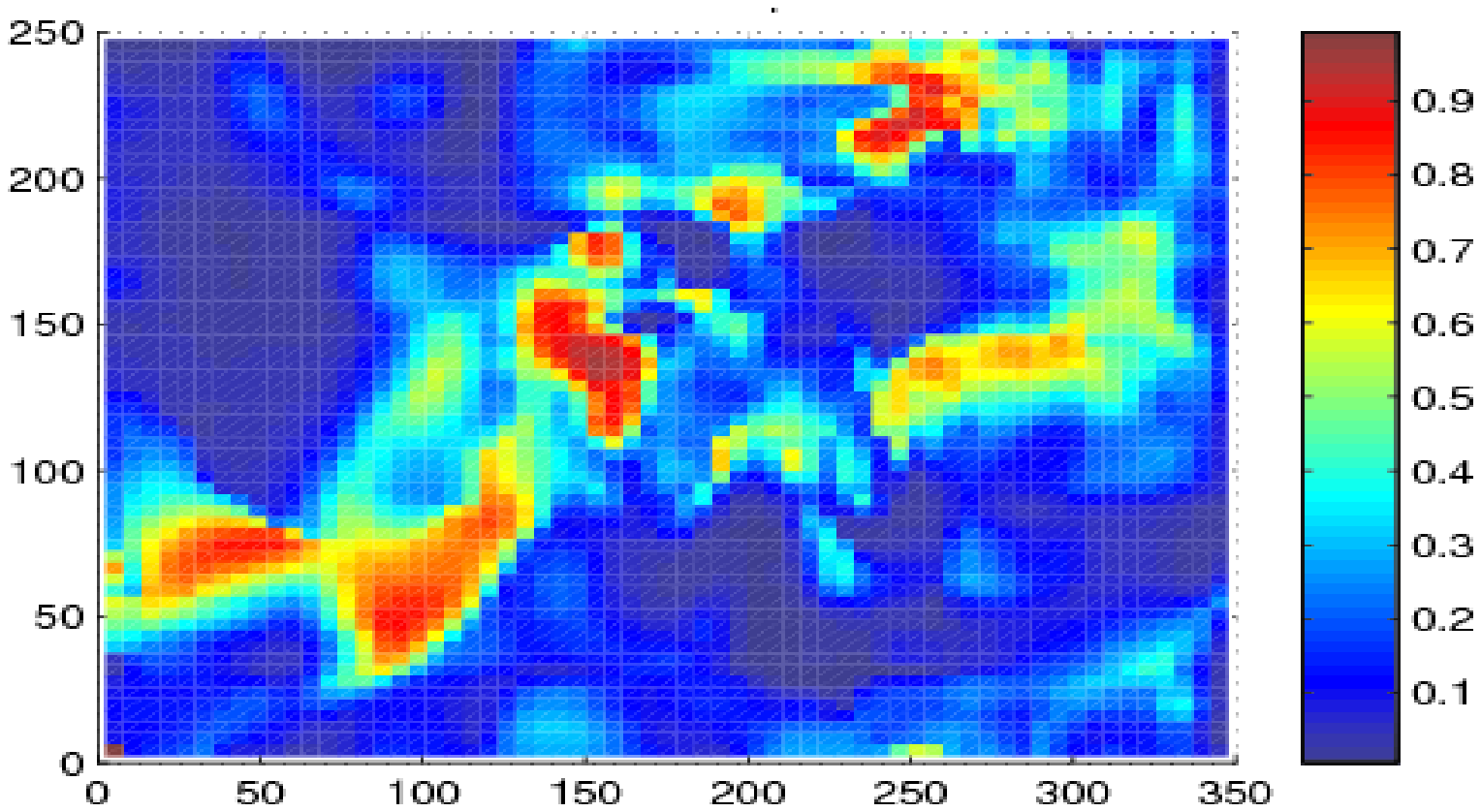,height=5.cm,width=8.5cm}
\psfig{figure=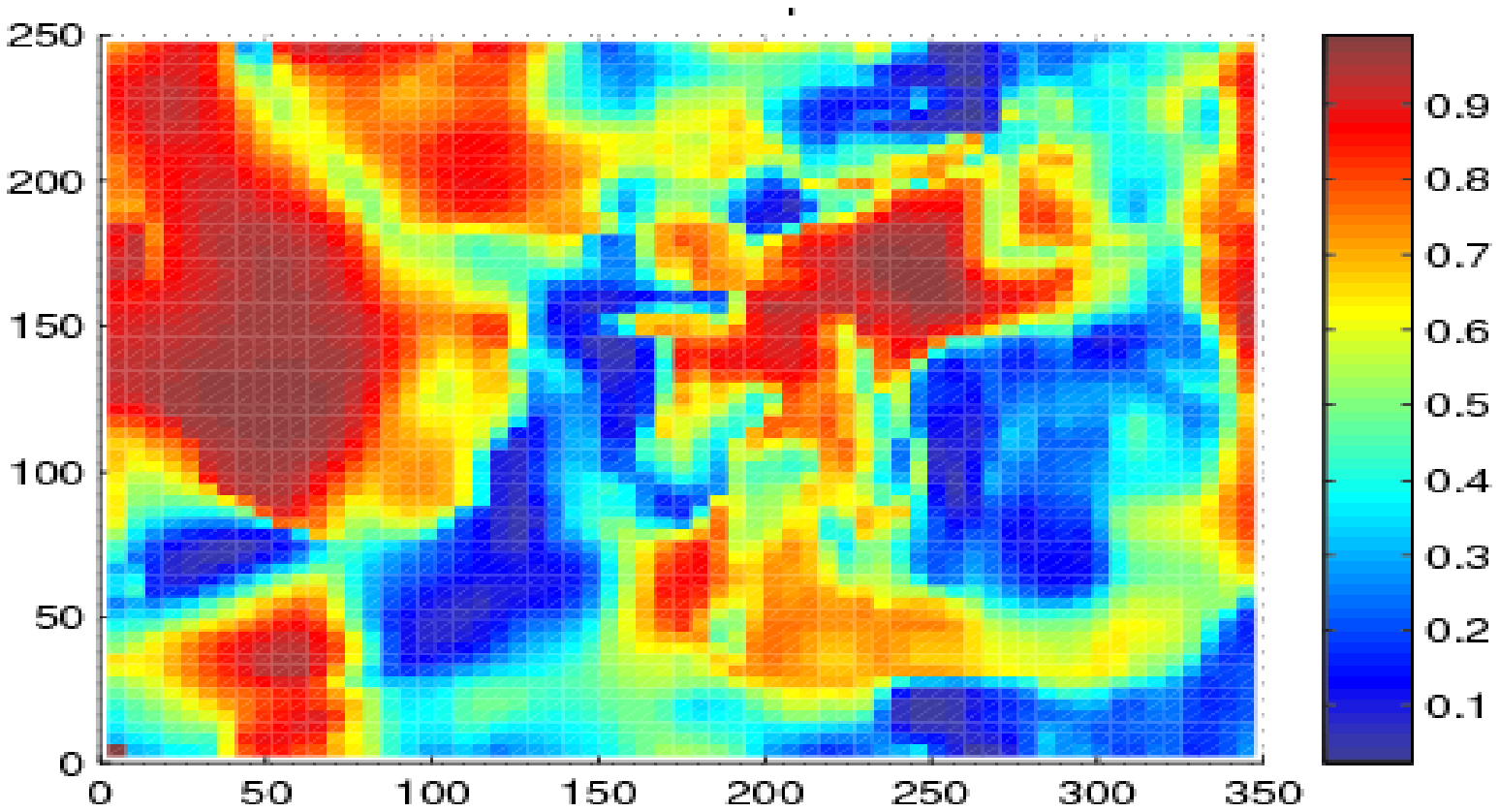,height=5.cm,width=8.5cm}$$
(a) \hspace{8cm} (b)
$$\psfig{figure=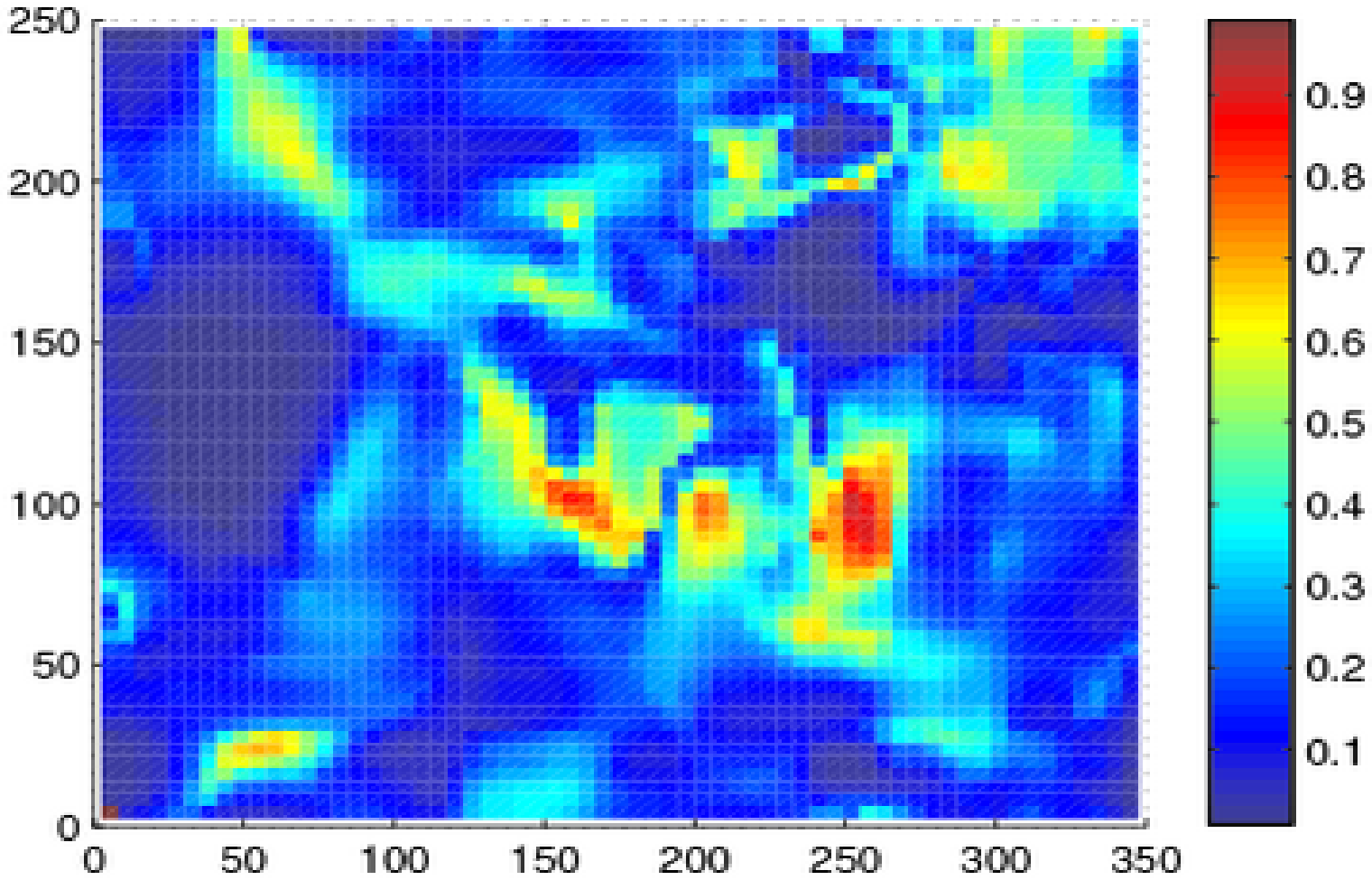,height=5.cm,width=8.5cm}
\psfig{figure=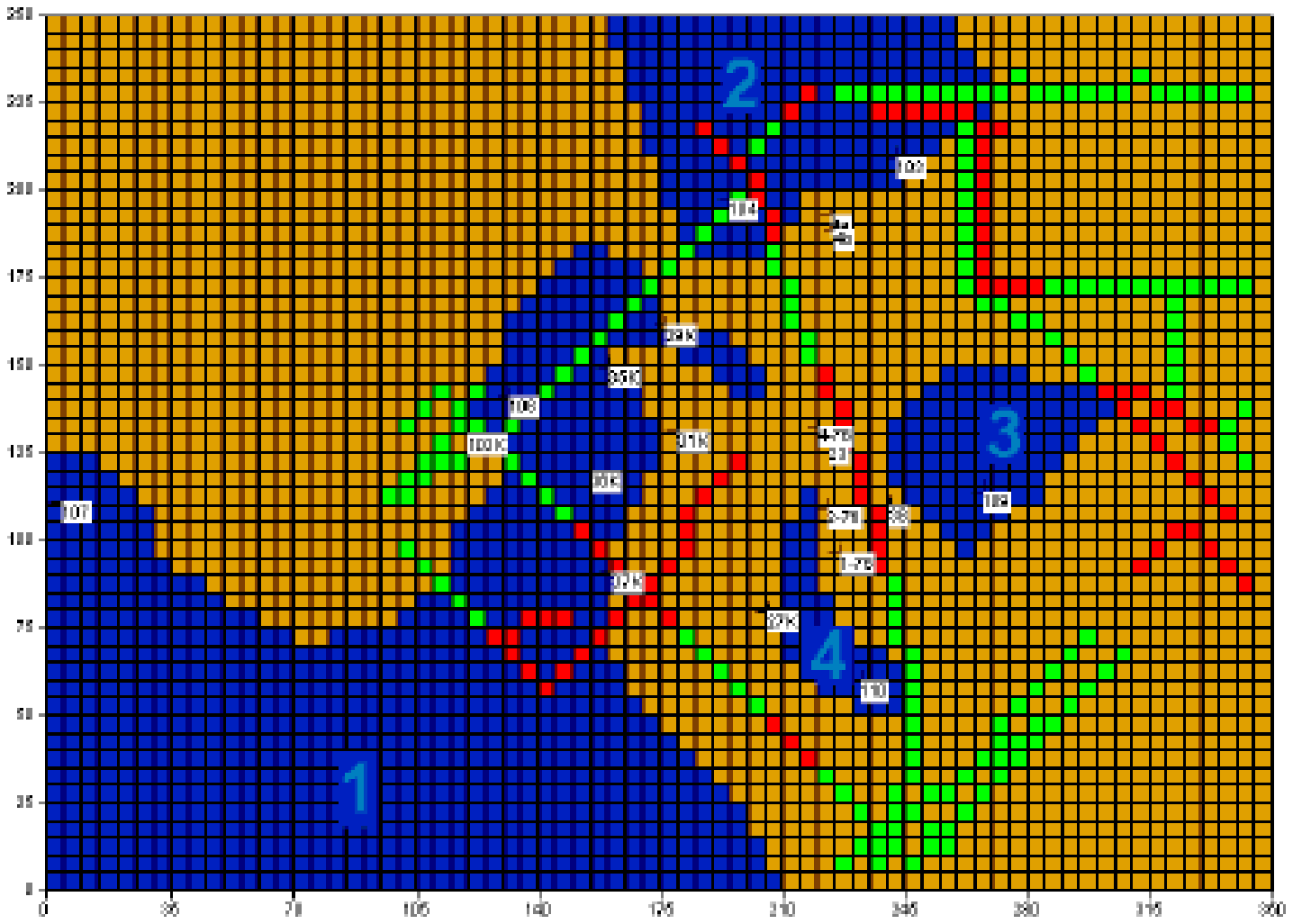,height=5.cm,width=7.5cm}
\hspace{1cm}$$
(c) \hspace{8cm} (d)
\end{center}

\caption{Total Sobol' indices of three input variables of MARTHE: (a) kd1 (distribution coefficient of the first layer), (b) kd2 (distribution coefficient of the second layer) and (c) i3 (high infiltration rate). (d): MARTHE hydrogeological model: blue zones (numbered from $1$ to $4$) correspond to low conductivity zones (absence of coarse sand in the second layer); lines present zones of high infiltration rates.}\label{fig:zones}
\end{figure}

Figure \ref{fig:zones} (a), (b) and (c) shows three maps of total Sobol' indices $S_{T_i}$ corresponding to the three main influential variables.
The $17$ remaining input variables have no influence in any zone of the spatial output domain.
These results are completely coherent with previous studies which have detected the predominant influence of these three variables.
Our new results have provided some additional spatial information.
For example, we locate more precisely the influence zones of the distribution coefficient of the first hydrogeological layer.
Such information is precious for model engineers.
It could help them to determine according to the spatial location of large variability zones the kind of additional information which is needed.
Subsequent decisions could be to place new piezometers in specific geographical zones.
The methodological developments highlight not only the direct application to post-treatment processes but also enable us to propose a new characterization strategy.

Figure \ref{fig:zones} (d) gives spatial information about the MARTHE model.
It clarifies the obvious correlation between the MARTHE hydrogeological scenario and our obtained spatial maps of sensitivity indices: influential $kd1$ zones correspond to the absence of the second hydrogeological layer while influential $kd2$ zones correspond to its presence.
In Figure \ref{fig:zones} (c), we also retrieve the high infiltration lines of Figure \ref{fig:zones} (d) and see their spatial area of influence.

    In our radioactive waste problem, the Sobol' maps of each uncertain input parameter clearly provide guidance to a better understanding of the simulator forecast and can be used to reduce the response uncertainties most efficiently. For example, if we want to reduce the predicted concentration uncertainty at a specific point of the map, we analyze all the Sobol' maps and determine the most influential inputs at this point.
    Then, we can try to reduce the uncertainty of these inputs by additional measures.
    Moreover, spatial maps for sensitivity indices can reveal gradient of influence of uncertain parameters, linked to the physics of the phenomenon (e.g. influence of a parameter varying in function of the flow direction). The global influence of each input over the whole space can also be used to identify areas of influence and areas of non-influence of this input and can be linked, as for $kd1$ and $kd2$, to a map of a geological parameter. If we now consider the strong infiltration coefficient denoted as $i3$ and its sensitivity map, we can deduce that $i3$ is only influential around the pipe and its influence is very limited outside the pipe area. 
The lack of knowledge on this parameter does not induce a big uncertainty on the concentration forecast at the site boundary and consequently on the decision relative to the need of a site rehabilitation.
    
\section{CONCLUSION}


In this paper, a new methodology was introduced to compute spatial maps of variance-based sensitivity indices (such as the Sobol' indices) for numerical models giving spatial maps as outputs.
Such situations often occur in environmental modeling problems.
One critical issue with our method is due to the reduced number of model output maps available because of the high cpu time cost of the numerical model.
A functional basis decomposition (wavelet basis) linked to a metamodel technique (based upon the Gp model) is proposed and used to solve this problem.
Choosing a wavelet basis is well-suited for our application cases (analytical and real models) because strong spatial heterogeneities and sharp boundaries are observed in the model output maps.
In addition,the Gp model is appropriate for handling the large differences between the output maps obtained for various inputs. This induces strong non-linear variations in the Gp-modeled wavelet coefficients.
The resulting functional metamodel is a fast emulator (i.e. with negligible cpu time) of the computer code.
It can be used for uncertainty propagation issues, optimization problems and, as advocated in this paper, for sensitivity index estimation.

An analytical test function was presented to explain the different steps, criteria and modeling choices of our methodology.
The convergence of our Gp-based functional metamodel was also investigated.
Then, our methodology was applied to a real case to stress its concrete applicability.
We particularly emphasized the relevance of the additional information (in addition to the expert and model knowledge) brought by the spatial maps of first order and total sensitivity indices.
These sensitivity maps allow us for spatially identifying the most influential inputs, for detecting zones with input interactions and for determining the zone of influence for each input.

Our methodology can be extended to any computer codes with functional outputs: codes with outputs depending on time, codes depending on other physical processes (such as a function of temperature), codes with outputs varying in space and time.
In the third case, the temporal and the spatial scales must be carefully distinguished.
It would be interesting in a future work to apply our method to the MARTHE spatio-temporal evolutions of the concentration values (between 2002 and 2010). In addition, improvements could be proposed.
For example, the vaguelette-wavelet decomposition \citep{abrsil97,ruiang07} would be an interesting substitute to the wavelet decomposition.
It would allow a simultaneous treatment of all the spatial output maps and a direct standardization of all decomposition coefficients.
Last, dealing with the functional input case remains an important and challenging issue to disseminate the global sensitivity analysis into environmental modeling communities.
\citet{ioorib08} and \citet{liltar09} proposed some preliminary methodologies to account for the spatially distributed inputs when computing Sobol' indices.


\section{ACKNOWLEDGMENTS}

This work was backed by the ``Risk Control'' project that is managed by the CEA/Nuclear Energy Division/Nuclear Development and Innovation Division, and by the ``Monitoring and Uncertainty'' project of IFP.
This work has also been backed by French National Research Agency (ANR) through COSINUS program (project COSTA BRAVA n$^o$ANR-09-COSI-015). 
We are grateful to Mickaele Le Ravalec for her help with the English.


\section*{APPENDIX A: SOBOL INDICES FOR THE CAMPBELL2D FUNCTION}\label{eq:app}

The analytical derivations of the first order Sobol' indices $S_i$ (Eq. (\ref{eq:indordre1})) of the Campbell2D function (\ref{eq:fct}) 
consists, first of all, in obtaining analytical expressions of the conditional expectations $\mathbb{E}\left(Y|X_i\right)$ (for $i=1,\ldots,8$).
The multiple integrations are made following the uniform distribution on $[-1,5]$ (we have $\mathbb{E}(X_i)=2$ and $\mbox{Var}(X_i)=3$ $\forall \;i=1,\ldots,8$).
The terms of these integrals which do not depend on $X_i$ can be directly put to zero (because these terms disappear when the variance over $X_i$ is taken).
In the next step, we take the variance over $X_i$ of the expressions of the conditional expectations (which leads to simple integrals).
In some cases, analytical simplifications can be made but in other cases, these variances cannot be simplified and the integrals are evaluated by Monte Carlo.

We recall that $(z_1,z_2) \in [-90,90]^2$ and we define the following variable changes:
\begin{equation}
\theta_1=0.8 z_1 +0.2 z_2 \;,\; \theta_2=0.5 z_1 +0.5 z_2 \;,\; \phi_1=0.4 z_1 +0.6 z_2 \;,\; \phi_2=0.3 z_1 +0.7 z_2 \;.
\end{equation}
The Campbell2D function is now written
\begin{equation}
	\begin{array}{r}
	g(\mathbf{X},z_1,z_2) = \displaystyle X_1\exp\left[-\frac{(\theta_1 -10X_2)^2}{60 X_1^2}\right]+(X_2+X_4)\exp\left[\frac{\theta_2 X_1}{500}\right] \\
	+ \displaystyle X_5 (X_3-2) \exp\left[-\frac{(\phi_1 -20 X_6)^2}{40 X_5^2}\right] +(X_6+X_8) \exp\left[\frac{\phi_2 X_7}{250}\right] \;,
	\end{array}
\end{equation}
We also define $\Phi(x)$ as the cumulative distribution function of a standardized Gaussian variable. 
The first order Sobol' indices for the $8$ input variables are written:

\begin{equation}\label{eq:C2DS1}
S_1(z_1,z_2) = \mbox{Var}\left\{\sqrt{\frac{\pi}{60}} X_1^2\left[\Phi\left(\frac{50-\theta_1}{\sqrt{30}X_1}\right)-\Phi\left(-\frac{10+\theta_1}{\sqrt{30}X_1}\right)\right]+4\exp\left(\frac{\theta_2 X_1}{500}\right)\right\} \;,
\end{equation}

\begin{equation}
S_2(z_1,z_2) =
\left\{\begin{array}{l} \displaystyle \mbox{Var}\left\{\frac{250X_2}{3\theta_2}\left[\exp\left(\frac{\theta_2}{100}\right)-\exp\left(-\frac{\theta_2}{500}\right)\right] + \int_{-1}^5 \frac{x}{6} \exp\left[-\frac{1}{2}\left(\frac{\theta_1-10X_2}{\sqrt{30}x}\right)^2\right]dx\right\}\\
\hfill \mbox{ if } \theta_2 \neq 0 \;,\\
\displaystyle \mbox{Var}\left\{X_2+\int_{-1}^5\frac{x}{6}\exp\left[-\frac{1}{2}\left(\frac{\theta_1-10X_2}{\sqrt{30}x}\right)^2 \right]dx\right\} \mbox{ if } \theta_2=0 \;,
\end{array}\right.
\end{equation}

\begin{equation}
S_3(z_1,z_2) = \frac{\pi}{120} \left\{ \int_{-1}^5 \frac{x^2}{6} \left[ \Phi\left(\frac{100-\phi_1}{\sqrt{20}x}\right)-\Phi\left(\frac{-20-\phi_1}{\sqrt{20}x}\right)\right]dx\right\}^2 \;,
\end{equation}

\begin{equation}
S_4(z_1,z_2) = 
\left\{\begin{array}{l} \displaystyle 
\frac{1}{3} \left\{ \frac{250}{\theta_2}\left[\exp\left(\frac{\theta_2}{100}\right)-\exp\left(-\frac{\theta_2}{500}\right)\right]\right\}^2 \mbox{ if } \theta_2 \neq 0 \;,\\
\displaystyle 3  \mbox{ if } \theta_2 = 0 \;,
\end{array}\right.
\end{equation}

\begin{equation}
S_5(z_1,z_2) = 0 \;, 
\end{equation}

\begin{equation}
S_6(z_1,z_2) = 
\left\{\begin{array}{l} \displaystyle 
\frac{1}{3} \left\{ \frac{125}{\phi_2}\left[\exp\left(\frac{\phi_2}{50}\right)-\exp\left(-\frac{\phi_2}{250}\right)\right]\right\}^2 \mbox{ if } \phi_2 \neq 0 \;,\\
\displaystyle 3  \mbox{ if } \phi_2 = 0 \;,
\end{array}\right.
\end{equation}

\begin{equation}
S_7(z_1,z_2) = 
\left\{\begin{array}{l} \displaystyle 
\frac{8}{3}\frac{125}{\phi_2} \left[\exp\left(\frac{\phi_2}{25}\right)-\exp\left(-\frac{\phi_2}{125}\right)\right]-\frac{4}{9}\left\{\frac{250}{\phi_2} \left[\exp\left(\frac{\phi_2}{50}\right)-\exp\left(-\frac{\phi_2}{250}\right)\right]\right\}^2 \\
\hfill \mbox{ if } \phi_2 \neq 0 \;,\\
\displaystyle 0  \mbox{ if } \phi_2 = 0 \;,
\end{array}\right.
\end{equation}

\begin{equation}\label{eq:C2DS8}
S_8(z_1,z_2) = S_6(z_1,z_2) \;.
\end{equation}

\singlespacing
\bibliographystyle{apalike}

\end{document}